\documentclass{aa}
\usepackage{epsfig}
\usepackage{rotating}
\usepackage{txfonts}
\usepackage{aalongtable}
\newcommand{\gesssim}{\mathrel{\hbox{\rlap{\hbox{\lower4pt\hbox{$\sim$}}}\hbox{$>$}}}}

\newcommand{\teff}{T$_{\rm eff}$}
\newcommand{\nli}{$\log n$(Li)}

\newcommand{\msun}{$\mathrm{M_{\odot}}$~}
\begin{document}

\title{An assessment of Li abundances in weak-lined and classical T Tauri stars of the Taurus-Auriga
association.\thanks{}}

   \subtitle{}

   \author{P. Sestito\inst{1} \and F. Palla\inst{1} \and S. Randich\inst{1}}

   \offprints{F. Palla}

\institute{INAF-Osservatorio Astrofisico di Arcetri, Largo E.~Fermi 5,
             I-50125 Firenze, Italy\\
\email{sestito, palla, randich@arcetri.astro.it}
            }

\titlerunning{Li abundances in low-mass stars of Taurus-Auriga}
\date{Received Date: Accepted Date}

  \abstract
 {Accurate measurements of lithium abundances in young low-mass stars provide
  an independent and reliable age diagnostics. 
  Previous studies of nearby star forming regions have identified
  significant numbers of Li-depleted stars, often at levels 
  inconsistent with the ages indicated by their luminosity.
   } 
{We aim at a new and accurate analysis of
Li abundances in a sample of $\sim$100 pre-main sequence
stars in Taurus-Auriga using a homogeneous and updated set of stellar 
parameters and model atmospheres appropriate for the spectral types of
the sample stars.
}
{We compute Li abundances using published values of the equivalent widths
of the Li $\lambda$6708~\AA~doublet obtained from 
medium/high resolution spectra.
}
{We find that the number of significantly Li-depleted stars in Taurus-Auriga
is greatly reduced with respect to earlier results. Only 13 stars 
have abundances
lower than the interstellar value by a factor of 5 or greater. 
All of them are weak-lined T Tauri stars drawn from X-ray surveys; 
with the exception
of four stars located near the L1551 and L1489 dark clouds, all the 
Li-depleted stars belong to the class
of dispersed low-mass stars, distributed around the main
sites of current star formation. If located at the distance of Taurus-Auriga,
the stellar ages implied by the derived
Li abundances are in the range 3-30 Myr, greater than the bulk of the Li-rich 
population with implication on the star formation history of the region.
}
{In order to derive firm conclusions about the fraction of Li-depleted
stars of 
Taurus-Auriga, Li measurements of the remaining members of the
association should be
obtained, in particular of the group of  stars that fall in the Li-burning 
region of the HR diagram.}

\maketitle

\section{Introduction}\label{intro}
The use of lithium as an alternative and secure method
to derive the ages of young, very low-mass (M$\sim$0.1--0.6~\msun)
stars is now well established on theoretical
and observational grounds. These stars are fully convective; thus they
have a simple internal structure and the physics of their Li depletion
during the pre-main sequence (PMS) phases
is well understood and not affected by major uncertainties.
Theory predicts that they start depleting their surface
Li after $\sim$2~Myr from  birth, and eventually exhaust all the internal
supply in about 10--50~Myr (e.g. D'Antona \& Mazzitelli~\cite{dantona94};
Baraffe et al. \cite{baraffe98}; Siess et al. \cite{siess}). 
The timescale of Li depletion depends on mass, with
lower mass stars being the slowest ones.
Therefore, in a cluster or association of a given
age there is a clear/sharp transition, the so-called lithium depletion
boundary (LDB), between
fully depleted objects and those with the initial lithium content. The mass
at which the boundary occurs reveals the age of the cluster.
Observationally, the LDB
test has been successfully applied to a number of young open clusters
in the age range 30-120~Myr (e.g.,
Stauffer et al.~\cite{stauf98}, \cite{stauf99};
Barrado y Navascu\'es et al.~\cite{byn};
Jeffries \& Oliveira~\cite{jo05}; Manzi et al.~\cite{manzi08}).

The physics of lithium depletion is considerably 
more complex for higher mass stars
in the range M$\sim$0.6--1.2~\msun that develop a radiative core during
PMS evolution. The exact amount of depletion in these stars depends on
a number of parameters and assumptions entering the models (e.g., Jeffries
\cite{jeff06}); hence, use of Li abundances to derive exact ages
of stars in this mass range is considerably uncertain, although it still allows 
distinguishing between very young and more evolved PMS stars.

In addition to providing absolute age estimates, Li abundances can
also be used to obtain information on the magnitude of the age spread within
less evolved system that are too young for the development of the LDB.
Recently, Palla et al. (\cite{pallaONC05}, \cite{pallaONC07}) have
measured Li abundances in a large sample of
low-mass members of the Orion Nebula Cluster (ONC), whose bulk population
is only $\sim$1--2 Myr old. They found that, in addition to the overwhelming
population of stars with Li abundances consistent with the initial
interstellar value (log N(Li)=3.1$\pm$0.2, e.g. Jeffries~\cite{jeff06}
and references therein), 
six high probability
members are strongly Li depleted, with Li depletion ages greater 
than 10 Myr. For four of these stars the agreement with the isochronal ages 
derived from the position of stars in the HR diagram is excellent. 
Similarly, Li depleted stars were found 
in Upper Scorpius (Mart\'{\i}n \cite{martin98}) and 
$\sigma$~Orionis (Sacco et al. \cite{sacco}). The presence of older stars
mixed with younger members reveals an extended phase of star formation that
lasted for much more than a few dynamical times.

Another region that has been the object of extensive Li studies in the past is
the Taurus-Auriga (Tau-Aur) association. The
largest collection of Li data has been presented in a series
of papers by
Basri et al. (1991, hereafter B91), Magazz\`u et al. (1992, hereafter M92),
 and Mart\'{\i}n et al. (1994, hereafter M94). These studies demonstrated for
the first time that the majority of young stars have  Li
abundances consistent with the initial 
interstellar value, but that a significant
number (about 30) stars are characterized by low or very low abundances. If
real, the observed depletion level would imply large  Li depletion ages 
($\gesssim$10~Myr), in excess of the bulk of the undepleted population 
($\gesssim$1--3~Myr). However, the interpretation of the data
remained controversial, since the measured Li abundances were not entirely
consistent with the predictions of theoretical models. In particular, the
lowest mass stars of the samples indicated too much Li destruction at the
observed luminosity. In other words, these T Tauri stars were too young, as
indicated by the isochronal ages, to have undergone significant nuclear
burning and Li destruction. The interpretation of this result was not 
straightforward and the possible explanations invoked 
large uncertainties in the stellar
parameters (effective temperature, luminosity, surface gravity), problems with
model atmosphere used to derive Li abundances, and limitations of the PMS 
evolutionary models. More recently, Wichmann et al. (2000; hereafter W00) 
have presented a detailed analysis of the young population discovered by
ROSAT in and near the Tau-Aur complex. Using high resolution echelle 
spectra, they measured lithium equivalent widths and used these measurements 
as a selection criterion for youth, expanding in this way the sample of
candidate low-mass TTS of Tau-Aur. 
Since the knowledge of the properties of the stellar 
population of Tau-Aur has greatly increased in the last decade (e.g.,
Kenyon \&  Hartmann 1995; Brice\~no et al. 2002; Luhman et al. 2003;
G\"udel et al. 2007), and new
atmospheric (e.g., Pavlenko et al. 2001) and evolutionary (Siess et al. 2000; 
Baraffe et al. 2002; D'Antona \& Montalb\`an 2003; 
Montalb\`an \& D'Antona 2006) models
have been developed, the time is ripe for a careful re-analysis of the Li 
abundances in T Tauri stars in this important star forming region.

The structure of the paper is the following: In Sect.~2 we describe the
sample, stellar parameters,  method of analysis, and results obtained by B91,
M92, M94 and W00; in Sect.~3 we present our new analysis of the sample and the
resulting  Li abundances based on updated stellar parameters and curves of
growth (COGs). 
In Sect.~4 we address issues related to the origin of the Li-depleted stars,
to the completeness of the sample and
the possibility of a new survey aimed at discovering new Li depleted TTS.
Sect.~5 closes the paper.

\section{Lithium in Taurus-Auriga:  previous observations}\label{literature}

Our new analysis of the Li abundance of TTS in Tau-Aur is based on the
results of the three major studies performed by B91, M92,
M94 and on that by W00.
Beside these studies, there have 
been other measurements of Li abundance in Tau-Aur in a small number of stars,
addressing specific issues such as the influence of accretion on the Li line
(e.g., Stout-Batalha et al. 2000) or the presence of accretion disks around
Li-depleted stars (e.g., White \& Hillenbrand 2005). Here, we summarize the 
observations and results obtained by B91, M92, M94, and W00.


Basri et al. (\cite{basri}):~the sample includes 49 classical and weak-lined T Tauri
stars (CTTS and WTTS, respectively) of Tau-Aur. 28 of these were
observed with the Hamilton echelle spectrometer at Lick Observatory, 
with a spectral resolution R$\sim 40,0000$; the remaining
21 stars were retrieved from
Walter et al. (1988) and Strom et al. (1989) and observed with
resolving powers between $10,000$ and $24,000$. Spectral-types of the whole
sample are in the range G2-M0. For the estimate of effective temperatures
(\teff), B91 adopted  the conversion
from spectral types by Cohen \& Kuhi (\cite{CK79}), while the luminosity was
taken from the literature. When necessary, measured Li equivalent widths
(EW) of the 28 stars observed at Lick
were corrected for veiling and the Li abundance was then derived using a
code developed by M. Spite (see Spite \& Spite \cite{plateau}) and model
atmospheres by Gustafsson (1982, private communication). Corrections for
non-LTE effects were not taken into account in the analysis. The main results
found by B91 are the following: (i) Li undepleted stars
have an average Li
abundance of \nli=3.6$\pm$0.3, larger (but consistent within the
uncertainties) than the initial interstellar value
and the maximum abundance
in young main-sequence F and G stars (\nli=3.1$\pm$0.2, 
Randich et al.~\cite{R01});
(ii) about 1/3 of the stars show large amount of Li depletion, 
much larger than predicted by models; most of these stars are cooler
than 4000~K; (iii) the Li pattern of CTTS and WTTS
is indistinguishable. 
B91 discuss the large uncertainties in the
stellar parameters and model atmospheres appropriate for TTS and conclude
that they represent a fundamental limitation for the determination of
accurate and reliable Li abundances.

Magazz\`u et al. (\cite{magazzu}):~M92 analyzed a sample of 13 stars in
Tau-Aur; for three of them (BP Tau, V410 Tau, GM Aur) they obtained 
new spectra with the Intermediate Dispersion Spectrograph (IDS) at
Isaac Newton Telescope (INT), providing a resolution R$\sim 20,000$. The
remaining stars were retrieved from Walter et al. (\cite{walter88}) and
are in common with the sample of B91. Note that, while B91
considered for their analysis the maximum Li EW given by Walter et al.,
M92 instead took an average of the minimum and maximum values.
To estimate \teff, M92 used the
calibration by de Jager \& Nieuwenhuijzen (\cite{dejager}), while the
luminosities were adopted from the literature. 
M92 performed veiling corrections for BP Tau and GM Aur with the
same technique of B91, but employed a different method for the determination of
Li abundances. Specifically, they 
extrapolated Gustafsson models atmospheres to higher atmospheric
levels, where the optical depth of the Li line is below 0.05,
and solved  the non-LTE problem for the Li
line in the \nli~range covered by their stars. COGs 
were then computed in both NLTE and LTE in the \teff~interval 4000-5000~K
and extrapolated for lower temperatures. For all but one Taurus members
the non-LTE abundances resulted higher
than the LTE value by $\sim$0.3--0.6 dex. In spite of these upward
correction, M92 also found examples of depleted stars (factors
up to $\sim 100$). Unlike B91, the
average interstellar abundance was found to be \nli=3.2$\pm$0.3, consistent 
with the meteoritic value. 

Mart\'{\i}n et al. (\cite{martin}):~M94 carried out new observations
of 32 CTTS and WTTS Taurus members/candidates using
the IDS spectrograph at the INT, ISIS
at the William Herschel telescope, and IACUB at the
Nordic Optical Telescope. 
The spectral resolution varied between
$R\sim10,000$ and 30,000. To this sample they added 23 stars retrieved
from the literature, for a total of 55 stars. Three stars were then discarded
from the sample due to their too strong H$\alpha$ emission and
veiling, while abundances for another two were not given
because they were too cold.
As M92, they adopted the temperature scale of de Jager \& Nieuwenhuijzen
(1987) taking into account both spectral-type and luminosity. Luminosities
for most of the stars were retrieved from the literature, while for
a few stars were calculated by M94 using their own photometry.
Since the sample included
only WTTS, M94 did not need to correct measured EWs for veiling effects.
Li abundances were computed both in LTE and non-LTE using the same
extrapolation method of M92, but model atmospheres by Kurucz,
rather than those of
Gustaffson. According to M94, their COGs
were more accurate than those of M92; they found
that NLTE
corrections are typically below $\sim$+0.1 dex, much
smaller than the error bars due to the uncertainty in the stellar parameters
($\sim$0.2--0.3~dex). Overall, M92 found an average value for the 
sample of \nli~3.1--3.2, thus confirming the correspondence with the
meteoritic value and settling the problem of the high abundance initially
determined by B91. The large majority of the TTS showed Li abundances in the
range 2.9--3.4, with a tail in the distribution of Li-depleted objects (15\%
of the total). In particular, all stars with luminosity log~L$<$0.5~L$_{\odot}$
were characterized by \nli$<$2.4, corresponding to depletion factors greater 
than 5. 

Wichmann et al. (\cite{wichmann00}):~W00 identified a sample of 22 
PMS candidate members of Tau-Aur,  
covering spectral range from G1 to M1.5. These stars were
selected on the basis of the X-ray properties, 
Li-equivalent widths, and proper motion. Spectra
were obtained with ELODIE at the Observatoire de Haute Provence with a 
resolution of $\sim 40,000$.
W00 measured Li EWs, but did not attempt to 
derive abundances on the assumption that the latter were uncertain and model
dependent. They also assigned spectral types based on the previous analysis of
Wichmann et al. (1996) and identified the PMS candidates by their position in
the effective temperature-equivalent width diagram and
comparison with the Pleiades distribution. 
Since in several cases the measured EWs 
were rather low (less than $\sim$200~m\AA), suggesting some level of Li
depletion, it is important to compute accurate Li abundances also for this 
group of WTTS.
\subsection{Previous results on lithium abundances}\label{lit2}
The three major studies by B91, M92, and M94 described above have allowed 
the determination of Li abundances in
76 stars of the Tau-Aur association, 21 CTTS and 55
WTTS. Their properties are listed in Table~\ref{Taurus} where for each star
we give a sequential number, name, class (c=classical, w=weak-lined), spectral
type, effective temperature, luminosity, surface gravity, EW(Li), Li
abundance from literature, reference, our newly derived Li abundances, and 
method used by us to derive \nli. Note that our determination of the Li 
abundance has been done using both the original values of the stellar 
parameters listed by the authors and our own most recent compilation 
(partially presented in G\"udel et
al. 2007; see below). This explains the multiple entries for each star given
in Table~\ref{Taurus}. In addition, 24 stars have been observed
more than once by different groups and all the values of EW(Li) and \nli~are
listed in the table: apart from few exceptions, the derived 
abundances are mutually consistent within the uncertainties.

Overall, the studies by B91, M92, and M94 have
proved that most TTS of Tau-Aur have Li abundances close to the
initial interstellar value. 
However, there are 40 measurements of the Li abundance in 31 stars below
a value of \nli=2.6~which we set (rather conservatively) as the threshold 
between depleted and undepleted stars, considering the uncertainty in the 
derived values. 
Of these 31 stars, 24 are WTTS and 7 CTTS (including DE~Tau, DF~Tau, DG~Tau, 
HL~Tau, UZ~Tau~W, GM~Aur, and DR~Tau). The amount of depletion varies 
between 3 and 
60 and in four cases only upper limits on \nli~ could be measured, indicating 
depletion factors much greater than 20. In four cases (035135$+$2528SE, 
Lk Ca 3, Lk Ca 4, and Lk Ca 7), the Li abundance derived by different authors 
for the same star is quite discrepant (up to a factor of 10), beyond the 
uncertainty of the individual measurements.

\begin{figure*}
\psfig{figure=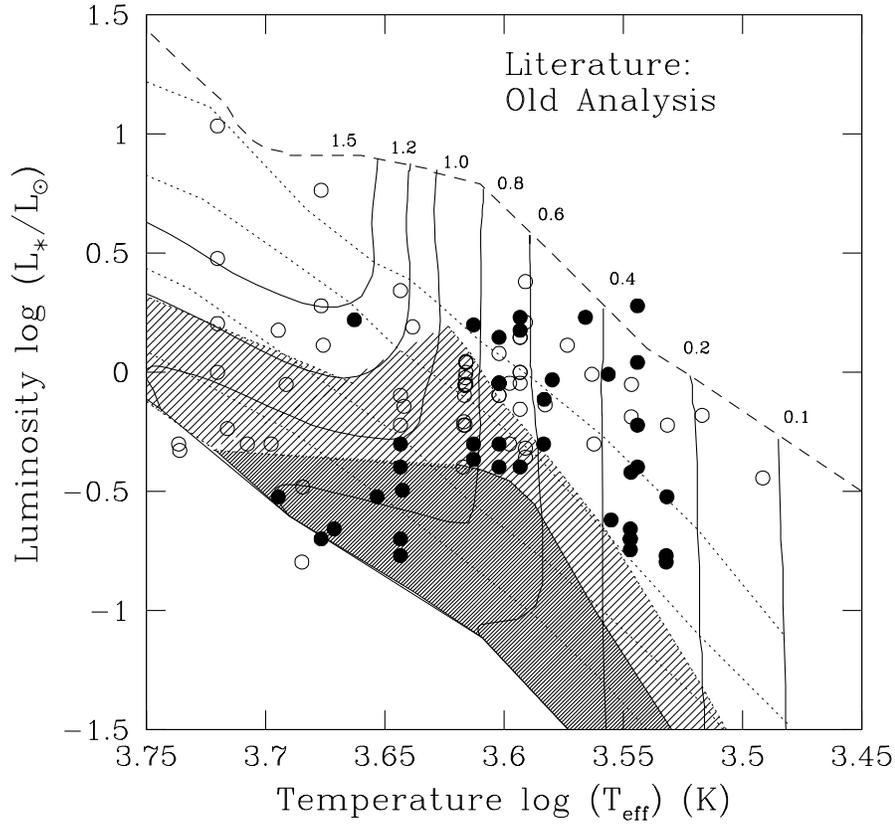, width=12cm, angle=0}
\caption{HR diagram for low-mass stars in Taurus Auriga
and location of the expected Li depletion
region. Evolutionary tracks for masses between 0.1 and 1.5 \msun~are indicated, 
as well as the birthline and isochrones of 1, 5, 10, and 30~Myr. 
The ZAMS is also shown.
The hatched regions indicate different levels of
predicted lithium depletion: \nli~down to
1/10-th of the initial value (light gray)
and below 1/10-th (dark gray). The stars were observed by B91,
M92, and M94,
and we adopted here stellar parameters and Li abundances
from the source papers (\nli~computed under LTE assumptions
by M92 and M94).
Filled circles represent Li-poor stars
(\nli$\leq$2.6), while the open ones are Li-rich stars.}\label{Li_litNLTE}
\end{figure*}

The distribution in the HR diagram of the 31 TTS stars with \nli$\leq$2.6 
is shown in Figure~\ref{Li_litNLTE}, along with the birthline,
evolutionary tracks and isochrones from Palla \& Stahler (1999). In addition,
Fig.~\ref{Li_litNLTE} displays the location of the predicted Li
depletion region, according to the models of Siess et al. (2000): the lightly
shaded gray region is for stars that have depleted up to 1/10th of the
initial Li content, while the dark gray zone is for depletion factors greater
than 10. In the figure, filled symbols indicate stars with \nli$\leq$2.6,
while open symbols are for stars with \nli$>$2.6.
Note that the 
distribution in the HR diagram would not change significantly by adopting NLTE
or non-LTE values. Among the 31 Li depleted stars, five of them are in the
region of expected partial depletion and five in that of complete depletion. 
Interestingly, the position of the five stars (034903$+$2431, 035135$+$2528NW,
035135$+$2528SE, 041529$+$1652, and 042835$+$1700) with depletion factors 
greater than 10 is consistent with the derived value of \nli~ (between 1.1 and
2.3). Two of these stars (034903$+$2431 and 042835$+$1700) have multiple 
values of stellar parameters and of \nli~ 
that place them also in the region of partial depletion, but close
to the border of complete destruction. The other three stars (VY Tau,
041559$+$1716, and IP Tau)
that fall in the light gray region have values of \nli~ consistent with the
theoretical prediction. The remaining 21 stars with 
\nli$\leq$2.6 are distributed in the portion of the HR diagram where
the luminosity is too high for any Li burning in the center. 
This fact highlights the basic problem of the interpretation of the survey 
by B91, M92 and M94, justifying a more refined assessment of both the stellar
parameters and lithium abundances, as we now discuss.
\section{Lithium in low-mass stars of Tau-Aur}\label{Our}
\subsection{New analysis of B91, M92, and M94}
Our new analysis of the available data consists of two steps. In the first
one, we recomputed the Li abundance using the input values for the stellar
parameters (T$_{\rm eff}$, L, and surface gravity) and EWs provided by the
authors. In the second step, we adopted the set of stellar parameters
recently compiled by one of us (F.P.) for the study of the overall population
of Tau-Aur and described in several papers (Palla \& Stahler 2000; Luhman et
al. 2003; G\"udel et al. 2007). The values of the stellar parameters are also 
listed for each star in Table~\ref{Taurus},
with the exception of two objects (035120$+$3154 and 045230$+$1746). We note that
the recent work by Bertout et al. (\cite{bertout07}) on the membership of a
subsample of TTS of Tau-Aur has 47 objects in common with our sample for
which they derived accurate estimates of the stellar parameters. In 
most cases their values agree with ours, but discrepancies in
spectral-type (and \teff) and/or luminosity are found for about 10 stars.
These objects will be further discussed in Sect.~\ref{bertout}.
Finally, for the calculation of 
derived quantities, such as mass (and thus surface
gravity), we used the PMS evolutionary models of Palla \&
Stahler (1999).

For the determination of the Li abundance, we have proceeded in different ways,
depending on the effective temperature: (a) for stars with \teff$\geq$3800 K,
we used both the spectral analysis code MOOG (Sneden 1973 --2002 version)
with model
atmospheres by Kurucz (\cite{kuru}) and (b) the COGs by Soderblom et al.
(\cite{S93}) or their extrapolation for stars with 
3800$\lesssim$\teff$\lesssim$4000 K; 
(c) stars cooler than 3800 K were 
analyzed with  COGs constructed by us and based on spectra provided by Ya.
Pavlenko (private communication; a partial set of COGs has been published by
Palla et al. 2007). The synthetic spectra cover a grid of parameters
from 3000 to 3800~K in \teff~(100~K steps), from 3 to 5 in $\log g$
(0.5~dex steps), and from $-1$ to 3.5 in \nli~(0.5~dex steps).
Different sets of COGs were computed depending on the
spectral resolution of the original data from each author: in particular,
R=40,000 for the B91 data, R=20,000 for M92, R=10,000/20,000/30,000 for M94.
Note that at the time when B91, M92, and M94 studies were performed,
researchers were referring to EWs rather than to pseudo-EWs (pEWs), i.e. equivalent widths of atomic
absorption lines measured with respect to the pseudocontinuum formed by the
haze of molecular lines (Pavlenko 1997). In our 
case, the COGs were derived based on measurement of pEWs on synthetic spectra
convolved to the different resolutions. Li
abundances were then derived from COGs (see Palla et al. 2007 for details)
under the assumption that the quoted EWs in the literature studies
were indeed pEWs. Note that, since synthetic spectra were available for
$-1 \leq n(Li) \leq 3.5$,
for several stars
cooler than 3800~K we were able to provide only a lower limit 
\nli$>$3.5 to Li abundances.

The resulting Li abundances and method used by us are given in the
last two columns of Table~\ref{Taurus}.
\subsection{Errors}
Errors in derived abundances are usually
due to uncertainties in stellar
parameters (with the major contribution coming from
errors in \teff) and EWs or pEWs. Given our approach, we do not consider
here uncertainties in stellar parameters, since abundances computed with the
different sets of stellar parameters already provide an idea of the systematic and random errors
involved. On the other hand, when the uncertainty in EWs
is available from the source paper, we estimated the corresponding error in 
$n(Li)$ and list it in Table~\ref{Taurus}.

As to other sources of systematic errors, we warn
that the COGs by Soderblom et al. (1993) were derived assuming
$\log g$=4.5 and  $\xi$=1.0 km s$^{-1}$, since they were optimized for main
sequence stars (F, G, and K-type). On the other hand 
in the MOOG analysis, we have adopted the $\log g$ values
listed in Table~1 and a microturbulent velocity $\xi$=1.5 km s$^{-1}$,
in agreement with literature estimates for PMS stars in the \teff~range 
$\sim 4000 -5000$~K (D'Orazi et a. \cite{dorazi}). For stars
with \teff~$\sim$4000~K up to 5000~K a change in $\xi$ of $\pm$0.5~km~s$^{-1}$
results in a difference of $\sim\pm$0.04--0.08 dex in \nli. Also,
differences of $\pm$1.0 dex in $\log g$ reflect into $\Delta$\nli~of
up to $\sim\mp$0.15 dex. Therefore, the possible errors introduced by using the 
COGs by Soderblom et al. (\cite{S93}) can be as large as $\sim$0.15--0.20~dex. 
For this reason, Li
abundances computed with MOOG are probably more reliable (particularly in
cases when $\log g$ significantly differs from 4.5), although the quoted
uncertainties are comparable to the errors in \nli~ and are not critical to
distinguish if a star is strongly Li depleted or not.
We note, however, that for stars with $\log g \sim 4.5$ the abundances derived
with MOOG are systematically higher than those derived with Soderblom's COGs 
and they would be even higher if assuming a $\xi=1.5$~km/sec.  While in most 
cases this offset does not affect our conclusions,
we will discuss in detail the few stars that turn out to be
Li depleted/undepleted if 
considering Soderblom or MOOG abundances, respectively.
\subsection{Li-abundance using the original stellar parameters}
\label{3.3}
\begin{figure*}
\psfig{figure=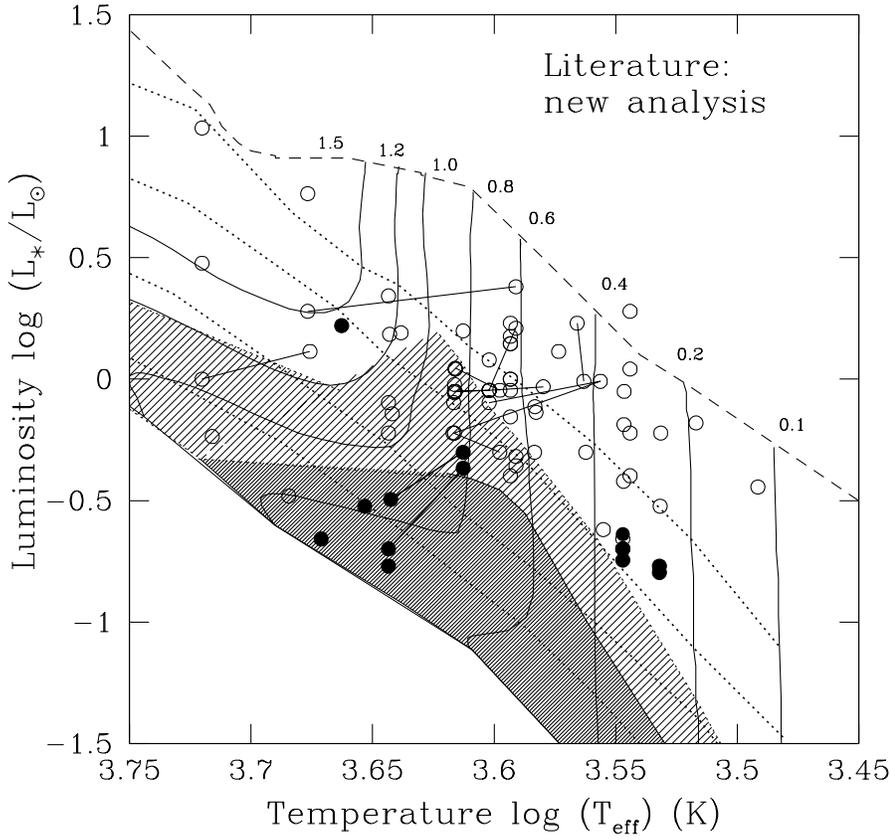, width=12cm, angle=0}
\caption{Distribution of WTTS and CTTS of Tau-Aur as in Fig.~1, but with 
Li abundances computed by us using stellar parameters from the source papers
and updated COGs.
Stars with multiple values of \nli~are connected by a line. 
Filled dots represent stars with \nli$\leq$2.6.
Note the large decrease of Li-depleted TTS compared to Fig.~1.}\label{Li_our}
\end{figure*}

Here we present the results of the calculation of \nli~ using as input values
the stellar parameters provided by the authors. The new \nli~ are listed
in the last two columns of Table~1 where we give the numerical value and the
method used to derive the abundance. For each star we also give a value
of the error on \nli~as derived from the uncertainty in the EW. It turns out
that the derived error is the same for abundances derived with MOOG and the COGs
by Soderblom. 
The main finding of the new analysis is that the values of \nli~ are 
systematically {\rm higher} than in the original derivations (discussed in 
Sect.~2.1). The average 
change is about $+$0.2--0.3 dex, but in a few cases differences by an order
of magnitude are found. Therefore, the net result is a substantial decrease of 
the number of Li-depleted stars. In particular, only 13 stars remain with
\nli~below the threshold value of 2.6. 
Interestingly, 
all of them are WTTs, but DR~Tau. Because of the difference between the 
abundances derived using MOOG or the COGs by Soderblom, the number of 
Li-depleted stars is 5 in the former case and 8 for the latter. Finally, the
5 coldest and least luminous TTS of the sample are confirmed as strongly 
Li-depleted using Pavlenko's analysis.
The distribution in the HR diagram of
the Li depleted and undepleted TTS is displayed in Figure~\ref{Li_our}.
Lines are used to connect the same star with different values of (L, \teff),
as given by the authors. Comparison with Fig. 1 immediately reveals that
the major inconsistency found in Sect. 2 is removed: 
all the stars in the upper portion of the diagram where no depletion is 
expected to occur have \nli~ consistent with the interstellar value.
In addition, several
TTS that lie close to the region of partial depletion in Fig. 1
are removed. On the other hand, the group of 5 stars of with spectral type 
M1-M2 and very low luminosity indeed remains Li-depleted at levels inconsistent
with the position in the HR diagram. Similarly, there is one early-type (K1)
and bright (log L$=$1.66 L$_\odot$) star, 042417$+$1744, with 
\nli$=$(2.62,2.32)$\pm$0.36 which is marginally Li-depleted, considering the
large uncertainty.
Finally, five stars in the dark-gray region are confirmed to be
Li depleted in the new analysis. Overall, 
these results suggest that the strong and unexpected Li depletion
found by B91, M92, and M94 for low-mass TTS is an artifact of
the analysis and the associated uncertainties.
\subsection{Li-abundance using updated stellar parameters}
\label{3.4}
In order to improve our analysis, we have examined the
individual values of luminosity and effective temperature of each star
using more recent compilations available in the literature and a homogeneous 
temperature scale for the conversion from spectral type to effective
temperature. In practice, we have used the data collected by one of us 
(F.P.) on the overall population of TTS in Tau-Aur and available in Palla
\& Stahler (2000) and G\"udel et al. (2007). The latter represents a
survey made with the XMM-Newton satellite of the X-ray properties of a 
large sample of TTS of Taurus, distributed along the major star forming 
sites (i.e. the filaments) of the molecular complex. G\"udel et al. (2007) 
provide comprehensive tables that summarize the stellar properties of all 
the target stars. In this way, we have been able to provide recent values 
of (L, \teff) for all the stars listed in Table~1. For the temperature scale,
we have used the conversion of Kenyon \& Hartmann (1995) for spectral types 
down to M0 and that of Luhman (1999; intermediate scale) for
later type stars. The new values are listed for each star in the last rows of
Table~1, along with the estimate of the surface gravity computed using the PMS
evolutionary models of Palla \& Stahler (1999). 

The main result of the exercise is the further reduction in the number of
reliable Li-depleted stars. We find only 9 stars (and 14 measurements) below 
the threshold value of \nli$=$2.6. Note that we do not consider V410 Tau
as depleted, since there is only one measurement below the threshold and it
is affected by a large uncertainty in the EW. Also, as mentioned, in two cases
(034903$+$2437, 035135$+$2528NW) the analysis from MOOG and Soderblom gives
discordant results, with the former slightly
above the threshold for Li-depletion. Considering this uncertainty, 
the new analysis shows that there are 
only 7 WTTs out of the original sample of 76 stars of Tau-Aur with secure
indication of severe Li-depletion. These
stars are: 040047$+$2603W, 040047$+$2603E, 040142$+$2150SW, 040142$+$2150NE,
040234$+$2143, 041529$+$1652, and 042835$+$1700. The other four Li-depleted 
stars according to the analysis of the previous subsection are now removed from
the list for the following reasons: 035135$+$2528SE, 042147$+$1744, and 
DR~Tau had
an effective temperature too low (4500~K, 4600~K, and 3920~K, respectively) for 
their spectral types (K0$=$5250~K, K1$=$5080~K, K7$=$4060~K); the luminosity,
and thus the surface gravity, 
of Anon 1 has changed
from 0.77 L$_\odot$ to 2.60 L$_\odot$.

\begin{figure*}
\psfig{figure=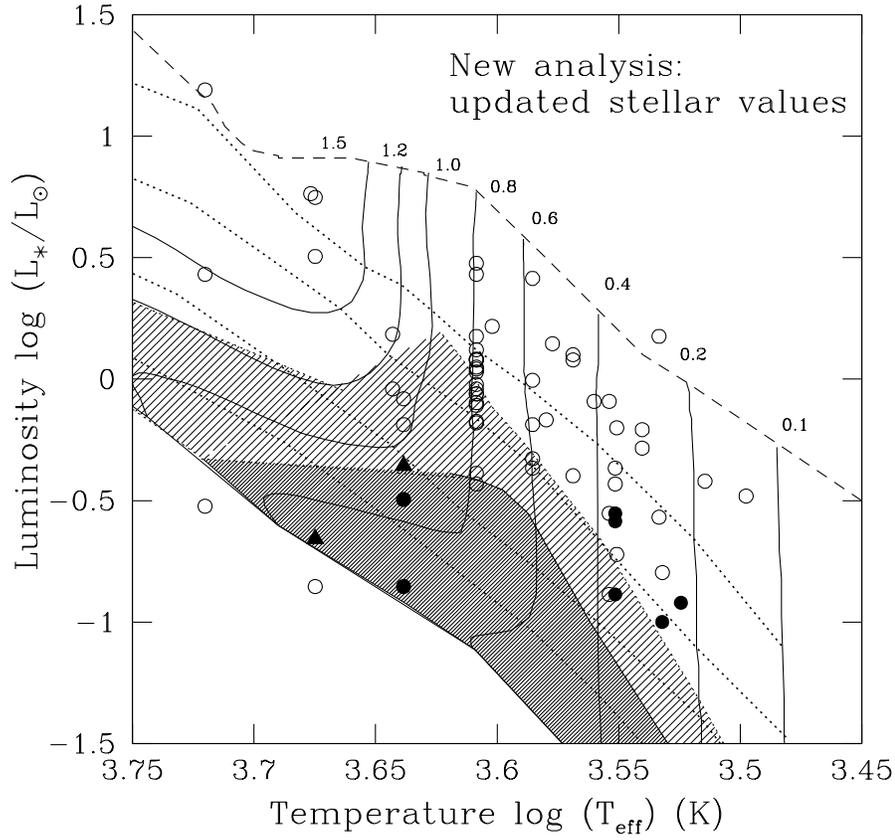, width=12cm, angle=0}
\caption{As in Fig.2, but with updated stellar properties. Filled circles refer
to bona fide Li-depleted stars, while triangles are for marginally Li-depleted
objects. Note the further improvement in the location of the Li-depleted 
and undepleted (empty circles) stars.}
\label{Li_ourXEST}
\end{figure*}

The new distribution of all the 76 TTS in the HR diagram is shown in 
Figure~\ref{Li_ourXEST}.
Only three stars fall the in region of full depletion, 2 in the region of
partial depletion, and 4 are outside (but close to the boundary of) the 
locus of Li-burning. The Li-depletion factor varies between $\sim$5 and 
$\gg$1000. As seen in Fig. 3, the
Li-depleted stars are divided in two groups: one with spectral types K3-K5 
falling in the fully depleted region, and one with late types (M2-M3) close
to the partial depletion region. For the first group, the observed Li-abundance
is consistent with the theoretical prediction for solar-type stars, with the 
exception of 035135$+$2528NW, that should be much more depleted for its
luminosity. This result, however, is not completely unexpected:
on the one hand, as mentioned in Sect.~1, 
the predicted amount of Li depletion for stars more massive than
$\sim 0.6$~\msun~is model dependent; on the other hand, we note that K-type
stars in the 30--50~Myr old clusters IC2602 and IC2391 are much less
Li-depleted than predicted by models (Randich et al.~\cite{R01}) and
have Li abundances \nli$\sim 2-2.5$, comparable to that of 035135$+$2528NW.
We conclude that, taken at face value, the implied isochronal ages of 
Li-depleted K-type stars
exceed 10 Myr and the Li ages might be even older, up to $\sim 30$ Myr: 
these stars might thus represent 
the tail of the much younger population of Tau-Aur.

The situation for the second group of stars is completely 
different, since
they show much larger amounts of Li-depletion for their luminosity.
These stars have estimated mass $\sim$0.2-0.4 M$_\odot$ and
thus should have fully convective interiors for which the theoretical 
predictions for Li-burning and depletion are quite robust and model 
independent (e.g., Bildsten et al. 1997).
In order to reconcile the Li-measurements with the stellar properties, the
latter should be incorrect by large factors (more than 200 K in temperature and
a factor of 3-4 in luminosity). Although large, such uncertainty is 
possible for these poorly studied WTTs.
\subsection{Li-abundance of the Wichmann et al. sample}
\begin{figure*}
\psfig{figure=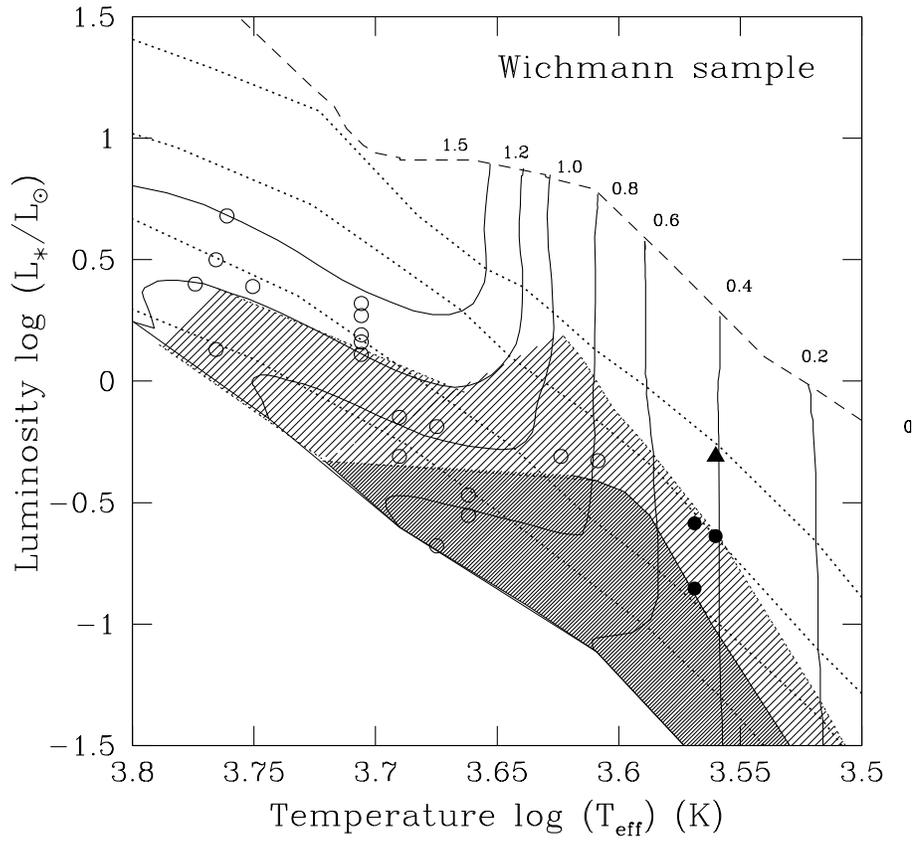, width=12cm, angle=0}
\caption{Location in the HR diagram of the 22 WTTS from Wichmann et al. (2000).
As in Fig.~3, the Li-depleted stars are shown by the filled symbols.}
\label{Li_wich}
\end{figure*}

We have computed the Li abundance for the 22 candidate PMS stars 
identified by W00 following the same method described in Sect.~3.1. 
Namely, for stars cooler than 3800~K we have used the COGs based on Pavlenko's
models, while for hotter stars we have employed both MOOG and the COGs by 
Soderblom et al. (1993). The stellar properties and derived \nli~ are listed in
Table~2. Spectral type, luminosity and EW(Li) have been taken from 
W00, while for the temperature scale and the derivation of surface
gravity we have proceeded as in Sect.~3.1.
\setcounter{table}{1}
\begin{table*}
\caption{Li abundances in T Tauri stars of Taurus-Auriga:
Wichmann et al. (2000) sample.}\label{Taurus2}
\begin{tabular}{lllllllll}
\hline
\hline
\# &Star & Sp.T.  & T$_{\rm eff}$& L & $\log g$ & EW &  n(Li) &  Method \\
&  RXJ & &   (K)        &  (L$_\odot$) &      & (\AA)&    &    \\
\hline
1 & 0403.3$+$1725 & K3 & 4730 & 0.65 & 4.38 & 0.49$\pm$0.03 &(3.97,3.71)$\pm$0.10 &a,b \\
2 & 0405.3$+$2009 & K1 & 5080 & 1.55 & 4.12 & 0.35$\pm$0.03 &(4.00,3.57)$\pm$0.18 &a,b \\
3 & 0405.7$+$2248 & G3 & 5830 & 3.16 & 4.08 & 0.25$\pm$0.03 &(4.31,3.72)$\pm$0.24 &a,b \\
4 & 0406.7$+$2018 & G1 & 5945 & 2.51 & 4.16 & 0.21$\pm$0.03 &(4.04,3.54)$\pm$0.21 &a,b \\
5 & 0409.1$+$2901 & K1 & 5080 & 1.29 & 4.17 & 0.41$\pm$0.03 &(4.31,3.93)$\pm$0.17 &a,b \\
6 & 0409.2$+$1716 & M1 & 3705 & 0.26 & 3.92 & 0.30$\pm$0.03 & 0.15$\pm$0.20 &c\\
7 & 0409.8$+$2446 & M1.5 & 3632 & 0.23 & 3.91 & 0.23$\pm$0.03 &$<$0.0 &c \\
8 & 0412.8$+$1937 & K6 & 4205 & 0.49 & 4.14 & 0.43$\pm$0.03 &(3.01,2.72)$\pm$0.11 &a,b \\
9 & 0420.3$+$3123 & K4 & 4590 & 0.28 & 4.49 & 0.37$\pm$0.03 &(3.24,2.97)$\pm$0.17 &a,b \\
10& 0423.7$+$1537 & K2 & 4900 & 0.71 & 4.30 & 0.36$\pm$0.03 &(3.72,3.38)$\pm$0.19 &a,b \\
11& 0424.8$+$2643 & K1 & 5080 & 2.09 & 4.04 & 0.41$\pm$0.03 &(4.34,3.93)$\pm$0.16 &a,b \\
12& 0435.9$+$2352 & M1.5 & 3632 & 0.49 & 3.54 & 0.49$\pm$0.03 &2.54$\pm$0.30 &c \\
13& 0437.2$+$3108 & K4 & 4590 & 0.34 & 4.41 & 0.36$\pm$0.03 &(3.19,2.91)$\pm$0.17 &a,b \\
14& 0438.2$+$2023 & K2 & 4900 & 0.49 & 4.51 & 0.32$\pm$0.03 &(3.49,3.12)$\pm$0.14 &a,b \\
15& 0438.2$+$2302 & M1 & 3705 & 0.14 & 4.21 & 0.19$\pm$0.03 &0.57$\pm$0.27 & \\
16& 0438.4$+$1543 & K3 & 4730 & 0.21 & 4.63 & 0.38$\pm$0.03 &(3.45,3.25)$\pm$0.16 &a,b \\
17& 0438.6$+$1546 & K1 & 5080 & 1.45 & 4.14 & 0.42$\pm$0.03 &(4.34,3.99)$\pm$0.14 &a,b \\
18& 0441.8$+$2658 & G7 & 5630 & 2.45 & 4.09 & 0.23$\pm$0.03 &(3.91,3.36)$\pm$0.22 &a,b \\
19& 0445.8$+$1556 & G5 & 5770 & 4.79 & 3.93 & 0.36$\pm$0.03 &(5.04,4.46)$\pm$0.20 &a,b \\
20& 0457.0$+$1517 & G3 & 5830 & 1.35 & 4.36 & 0.20$\pm$0.03 &(3.84,3.36)$\pm$0.20 &a,b \\
21& 0457.2$+$1524 & K1 & 5080 & 1.86 & 4.07 & 0.45$\pm$0.03 &(4.17,4.12)$\pm$0.10 &a,b \\
22& 0458.7$+$2046 & K7 & 4060 & 0.47 & 4.04 & 0.45$\pm$0.03 &(2.92,2.66)$\pm$0.10 &a,b \\
\hline		      	        	 	 	  
\hline		     	        		 
\end{tabular}	

Method -- a: MOOG; b: COGs by Soderblom et al. (1993); c: our own COGs. 
\end{table*}
The analysis shows that only four WTTS have Li abundances well 
below the initial value. These stars are the coolest of the sample and 
their position in the HR diagram is displayed in Figure~\ref{Li_wich}, along 
with the other WTTS. In three cases, the amount of Li-depletion is marginally
consistent with the predictions of theoretical models, while the fourth
star (RXJ0435.9$+$2352) that lies above the Li-depletion region has
a value of \nli$\simeq$2.6$\pm$0.3. In this respect, the 
situation is similar to that found for the five low-mass, Li-depleted stars
shown in Fig.~3. 
The fact that all the other stars of the W00 sample do not
show evidence for Li-depletion is not too surprising considering that 
the selection criterion used for the identification of the candidate TTS
was the presence of the Li absorption line with an equivalent width greater 
than 100~m\AA. What is more surprising is the high values of the derived \nli~ 
obtained with both methods, in most cases well in excess of the standard 
interstellar value. We explain this result as due to poor measurements
of the intrinsic EWs for stars that are rapid rotators. In fact, most of the 
stars listed in Table~2 have values of v$\sin i$ greater 
(or much greater) than 
10--20~km~s$^{-1}$. Viceversa, three of the Li-depleted stars are slow 
rotators (v$\sin i\sim$6-8~km~s$^{-1}$), hence their EW measurements are more
reliable. On the other hand, the fourth Li-depleted star, RXJ0409.2$+$1716,
is a fast rotator, but the nature of the system is uncertain (possibly an
SB2 system according to W00). 
\subsection{Comparison with the sample of Bertout et al. 2007}\label{bertout}
As anticipated in Sect.~3, Bertout et al. (2007) have studied a group of
about 70 stars of Tau-Aur with known parallaxes, belonging to a newly
identified moving group (see Genova \& Bertout 2006). For all of them, they
have re-derived photospheric luminosities, while most spectral-types and
effective temperatures were taken
from Kenyon \& Hartmann (1995). Overall, there are 47
stars in common with our sample, 43 of those listed in Table~1 and 4 from
W00. In most cases, the stellar parameters are very
similar to those adopted by us and the difference does not affect the
derivation of the Li abundance. Significantly, all of the Li depleted stars
in common with Bertout et al. (2007) have the same values of L and 
T$_{\rm eff}$.

However, there are 13 stars for which the discrepancy with our values
is quite large, beyond the error uncertainty in the determination of both 
luminosity and spectral types. To evaluate the impact on the determination of
the Li abundance, we have recomputed \nli~for all of them, assuming
\teff~values and luminosities given by Bertout et al. 
The resulting
values are listed in Table~3, with the usual notation for the abundances
derived with MOOG, Soderblom et al., and with our COGs. We find that even
with the new parameters the derived abundances are consistent with 
the initial interstellar value. Therefore, we conclude that
there are no additional Li depleted stars in addition to those already
discussed in the previous sections.
\setcounter{table}{2}
\begin{table*}
\caption{Stars with discrepant stellar parameters in Bertout et al. 2007.
n(Li)$_{\rm BSC}$ indicates Li abundances computed using stellar parameters
(\teff, L) of Bertout et al. 
}\label{Taurus3}
\begin{tabular}{lllllllll}
\hline
\hline
\# &Star & Sp.T.  & T$_{\rm eff}$& L & $\log g$ & EW &  n(Li)$_{\rm BSC}$ &  
Method \\
&  & &   (K)        &  (L$_\odot$) &      & (\AA)&    &    \\
\hline
1 & Hubble~4           & K7 & 4060 & 0.48 & 4.05 & 0.61 & 3.33,3.08 & a,b \\
2 & V410~Tau~ABC       & K4 & 4730 & 1.41 & 4.07 & 0.42 & 3.83,3.45 & a,b \\
  &                    &    &      &      &      & 0.57 & 4.33,3.96 & a,b \\
3 & RY~Tau             & K1 & 5080 & 6.59 & 3.72 & 0.27 & 3.48,2.99 & a,b \\
4 & IP~Tau             & M0 & 3850 & 0.34 & 3.98 & 0.50 & 2.87,2.65 & a,b \\
5 & V927~Tau~AB        & M3 & 3470 & 0.60 & 3.25 & 0.51 & 3.52 & c \\
6 & 043124$+$1824      & G8 & 5520 & 2.93 & 4.03 & 0.31 & 4.42,3.84 & a,b \\
  &                    &    &      &      &      & 0.23 & 3,78,3,23 & a,b \\
7 & DR~Tau             & K7 & 4060 & 1.97 & 3.43 & 0.46 & 3.16,2.69 & a,b \\
8 & GM~Aur             & K3 & 4730 & 1.23 & 4.11 & 0.50 & 4.14,3.74 & a,b \\
  &                    &    &      &      &      & 0.59 & 4.38,4.02 & a,b \\
9 & RW~Aur~A           & K4 & 4590 & 1.72 & 3.95 & 0.67 & 4.36,3.88 & a,b \\
10 & RXJ~0405.7$+$2248 & G0 & 6030 & 5.47 & 3.95 & 0.25 & 4.53,3.93 & a,b \\
11 & RXJ~0406.7$+$2018 & F8 & 6200 & 3.78 & 4.12 & 0.21 & 4.29,3.78 & a,b \\
12 & RXJ~0423.7$+$1537 & G5 & 5770 & 1.00 & 4.46 & 0.36 & 4.89,4.46 & a,b \\
13 & RXJ~0457.0$+$1517 & G0 & 6030 & 1.75 & 4.34 & 0.20 & 4.01,3.55 & a,b \\
\hline		      	        	 	 	  
\hline		     	        		 
\end{tabular}	

Method -- a: MOOG; b: COGs by Soderblom et al. (1993); c: our own method

\end{table*}
\section{Li-depleted stars of Tau-Aur}
Overall, our new determination of Li abundances has shown that of the about 
100 low-mass stars of Tau-Aur with Li EW measurements, only 13 objects 
can be considered bona-fide Li-depleted, i.e. with values of \nli~
lower than 2.6. A summary of the properties of these stars is given
in Table~\ref{summa}, where in addition to the quantities introduced
in Tables~1 and 2 we give the radial velocity (from Walter et al. 1988;
Hartmann et al. 1986; W00), the proper motion (from Doucourant et al. 2005),
and the kinematic membership to the Tau-Aur moving group (from Bertout \& 
Genova 2006).
All of the Li-depleted stars are WTTS drawn from X-ray selected surveys
({\it Einstein} for Walter et al. 1998 and {\it ROSAT All Sky Survey} for
Wichmann et al. 2000) and are
widely dispersed across the complex, away from the main star forming sites. 
The few exceptions are 041529$+$1652 and 042835$+$1700 in the area of the
L1551 region; RXJ0435.9$+$2352 and RXJ0438.2$+$2302 in the L1536 region. 
\setcounter{table}{3}
\begin{table*}\scriptsize
\caption{Properties of the Li-depleted stars of Taurus-Auriga.}\label{summa}
\begin{tabular}{lllllllllcrrc}
\hline
\hline
\# &Star & Sp.T.  & T$_{\rm eff}$& L & $\log g$ & EW & n(Li) &  Method & 
V$_{\it rad}$ & \multicolumn{2}{c}{PM} & Memb.\\
&     &  &  & & & & & & & $\mu_\alpha cos \delta$ & $\mu_\delta$ & \\
&     &  &   (K)        &  (L$_\odot$) &          & (\AA)&    & & (km s$^{-1})$
& (mas/yr) & (mas/yr) & \\
\hline
1 & NTTS034903$+$2431 &K5&4350 &    0.44         &   4.30 &   0.37$\pm$0.08 &   (2.93,2.62)$\pm$0.20  & a,b & 9.8 & 20 & $-$48 & N \\
  &             &  &     & & &  0.33$\pm$0.02 &        (2.70,2.38)$\pm$0.14  & a,b & & & & \\
&& &    &   & &  0.31$\pm$0.02 &         (2.57,2.25)$\pm$0.14  & a,b & & & & \\
&& &    &   & &  0.35$\pm$0.02 &         (2.81,2.50)$\pm$0.13  & a,b  & & & &\\
& & & & & & & &  & & & & \\
2 &NTTS035135$+$2528NW&K3&4730&     0.2  &    4.60   &   0.22 & 2.51,2.17  & a,b
& 9.9 &-- & --& --\\
 &&&      &   & & 0.24$\pm$0.01 &  (2.65,2.30)$\pm$0.08  & a,b & & & & \\
 & & & & & & & & & & & & \\
3 &NTTS040047$+$2603W &M2&3560 &    0.28 &   3.78 &  $<$0.07&     $<-$0.30     & c & 14.4 & 27 & $-$36 & N \\ 
& & & & & & & & & & & & \\
4 &NTTS040047$+$2603E &M2&3560 &    0.26 &   3.88 &  $<$0.06&     $<-$0.50    & c & 14.4 & 6 & $-$23 & Y \\
& & & & & & & & & & & & \\
5 &NTTS040142$+$2150SW&M3.5&3345 &  0.12 &   3.78	&  $<$0.25&    $<$1.34  & c & 16.2 & $-$11 & 8 & N \\ 
& & & & & & & & & & & & \\
6 &NTTS040142$+$2150NE&M3&3415 &    $>$0.05 & $<$4.23 &  $<$0.30& $<$1.74  & c & 16.2 & $-$17 & 0 & N \\ 
& & & & & & & & & & & & \\
7 &NTTS040234$+$2143  &M2&3560 &    0.13 &   4.06 &  0.34$\pm$0.02   &    1.84$\pm$0.17       & c & 16.7 & 31 & 11 & N \\ 
& & & & & & & & & & & & \\
8 & RXJ0409.2$+$1716 & M1 & 3705 & 0.26 & 3.92 & 0.30$\pm$0.03 & 0.15$\pm$0.20 &c & 14.7 & 3 & $-$11 & Y \\
& & & & & & & & & & & &  \\
9 & RXJ0409.8$+$2446 & M1.5 & 3632 & 0.23 & 3.91 & 0.23$\pm$0.03 &$<$0.0 &c & 10.9 & 30 & $-$43 & N \\
& & & & & & & & & & & & \\
10 &NTTS041529$+$1652  &K5&4350 &    0.14 &   4.67 &  0.19$\pm$0.01   &    (1.77,1.46)$\pm$0.07  & a,b & 15.8 & 11 & $-$14 & Y \\
& & & & & & & & & & & & \\
11 &NTTS042835$+$1700 &K5&4350 &    0.32 &   4.39 &   0.15 &	   1.39,1.24  	& a,b & 16.7 & 10 & $-$28 & Y  \\
&&&  & & &  0.19$\pm$0.04 & 	    (1.72,1.46)$\pm$0.23	& a,b & & & & \\
&&&  & & &  0.11$\pm$0.01 & 	    (1.11,1.01)$\pm$0.08	& a,b & & & & \\
& & & & & & & & & & & & \\ 
12& RXJ0435.9$+$2352 & M1.5 & 3632 & 0.49 & 3.54 & 0.49$\pm$0.03 &2.54$\pm$0.30 &c & 16.9 & 13 & $-$23 & Y \\
& & & & & & & & & & & & \\
13& RXJ0438.2$+$2302 & M1 & 3705 & 0.14 & 4.21 & 0.19$\pm$0.03 &0.57$\pm$0.27 & c & 15.7 & $-$8 & $-$16 & N\\
\hline		      	        	 	 	  
\hline		     	        		 
\end{tabular}	

Method -- a: MOOG; b: COGs by Soderblom  et al.  (1993); c: our COGs 

\end{table*}
This raises the question on the origin of this small group of Li-depleted 
stars, and in particular whether they can be considered as members of Tau-Aur. 
In fact, we note that ten out of 13 stars have radial velocities between 14 
and 17~km~s$^{-1}$ (see Table~\ref{summa}), consistent 
with membership since the distribution of the overall population is strongly 
peaked at 16--17~km~s$^{-1}$ (Walter et al. 1988, Hartmann et al. 1986). 
Of these ten stars, five have proper motions (also listed in Table~\ref{summa}) 
consistent with membership according to the analysis by Bertout \& Genova
(2006) who identified a Tau-Aur moving group on kinematic grounds. 
Only three of them (040047$+$2603E, 041529$+$1652, 042835$+$1700) belong to 
the moving group, while there is no information on the other 
two stars (RXJ0409.2$+$1716 and RXJ0435.9$+$2352). Interestingly, the three 
kinematic members are located near the L1551 and L1489 dark clouds and
may represent the earliest stars that formed within these units. 

Several of the other seven stars that are non-kinematic members of the moving 
group according to the proper motion analysis have radial velocities consistent 
with membership to the Tau-Aur association and are located 
away from the current sites of star formation. 
Brice\~no et al. (1997) have argued that the majority of
the dispersed population of X-ray active stars is the result of different
star formation episodes in a variety of star forming sites that have spread
with time over large distances. However, they concluded that in addition
there could be a few, widely dispersed and relatively young PMS stars selected 
on the basis of their X-ray and Li properties that had been missed by
previous searches limited to the boundaries of the molecular cloud
complex. The small group of Li-depleted stars discussed here could be an 
example of such a population.

We conclude that the low Li abundance and the kinematic properties, 
along with the position in the HR diagram, confirm the presence of a 
subpopulation of stars more evolved than the bulk of the Li-undepleted stars,
but younger than the diffuse background population with ages 
$\lesssim$100~Myr.
\subsection{Completeness of the sample}\label{completeness}
The sample of $\sim$100 low-mass stars with accurate measurements of
EW(Li) and \nli~ represents about 1/3 of the known (candidate) members of 
Tau-Aur. According to our own compilation, the current tally includes about 
300 members and the number is bound to increase as soon as follow-up 
observations of new candidates identified by 
{\it XMM-Newton} (Scelsi et al. 2007) and {\it Spitzer} (Rebull et al. 2007)
will become available. Therefore, the conclusion of our abundance analysis
on the small number of Li-depleted stars, although statistically
significant, is still not conclusive. A rather large number of known members
without lithium EW measurements
do actually fall within or close to the Li-burning
region in the HR diagram. As shown in Fig.~5, considering only stars with mass
$\lesssim$0.6~M$_\odot$, there are more than 20 objects for which the 
determination of \nli~would be highly desirable.
\section{Conclusions}\label{conclusions}
We have computed Li abundance in a sample of $\sim$100 T Tauri stars of 
Tau-Aur with published values of EW(Li) obtained from high resolution spectra.
The Li abundance was derived using updated atmospheric models and a set
of homogeneous stellar data as inputs. The most significant results are the
following:
\begin{itemize}
\item 
The serious inconsistency found in previous studies on the relatively 
high number of Li-depleted of TTS at the wrong luminosity is removed. Only 
13 stars have Li abundance
below a value \nli$=$2.6, corresponding to depletion factors greater 
than five. 
We attribute the early findings to incorrect Li-abundance analysis
and/or to the use of inconsistent stellar parameters.
\item
The majority of the Li-depleted stars lie in the HR diagram either within or
close to the boundary of the theoretical region of Li-burning in low-mass
stars. Conversely, most of the remaining stars with Li abundance consistent
with the initial interstellar value are distributed in the appropriate
portion of the HR diagram. 
\item
Unlike previous results, none of the 21 CTTS included in the sample shows 
evidence for Li-depletion. On the other hand, all of the 13 Li-depleted stars
are WTTS identified in X-ray surveys. Most of them
belong to the widely dispersed
population distributed away from (or at the periphery of) the main star forming
sites. In several cases the radial velocity and proper motions are consistent
with membership to the Tau-Aur moving group.
\item
If located at the distance of Tau-Aur, the Li depletion and isochronal
ages of the Li-depleted stars are in the range $\gesssim$3--15~Myr, greater 
than the bulk of the undepleted population.
\item
In order to draw firm conclusions on the Li-content of TTS in Tau-Aur, it is
necessary to obtain high quality measurements of the remaining known 
members of the complex, which outnumber the group studied here by a 
factor $\sim$3. In particular, a group of $\sim$25 stars that lie inside the
Li-burning region constitute an interesting subsample that will be the
subject of a future study.
\end{itemize}
\begin{acknowledgements}
We acknowledge partial support from PRIN-INAF "Stellar clusters: test for 
stellar formation and evolution" (PI: F. Palla).
\end{acknowledgements}
{}
\setcounter{table}{0}
\begin{longtable}{llllllllllll}
\caption{Li abundances in T Tauri stars of Taurus-Auriga.}\label{Taurus}\\
\hline
\hline
\# &Star &Class& Sp.T.  & T$_{\rm eff}$& L & $\log g$ & EW & n(Li)$_{\rm lit}$& Ref. &  n(Li)$_{\rm our}$ &  Method \\
&     & &  &   (K)        &  (L$_\odot$) &          & (\AA)&                   &                    &      &  \\ \hline
\endfirsthead
\caption{continued.}\\
\hline
\hline
\# &Star &Class& Sp.T.  & T$_{\rm eff}$& L & $\log g$ & EW & n(Li)$_{\rm lit}$& Ref. &  n(Li)$_{\rm our}$ &  Method \\
&     & &  &   (K)        &  (L$_\odot$) &          & (\AA)&                   &                    &      &  \\ \hline
\endhead
\hline\endfoot
1 & 032641$+$2420 &w&K1&5100 &    0.5 &   4.5 &   0.47 & 4.3 & B91& 4.43,4.14  & a,b \\
 &  &&K1&4985 &    0.5 &   4.5 &   0.35$\pm$0.02 & 3.4$\pm$0.35 & M94& (3.72,3.37)$\pm$0.20  & a,b \\
 &  &&K1&5080 &    0.5 &   4.5 &   0.47 &  & us& 4.41,4.13  & a,b \\
 &  &&K1&5080 &    0.5 &   4.5 &   0.35$\pm$0.02 &  & us& (3.92,3.58)$\pm$0.12  & a,b \\
 & & & & & & & & & & & \\
2 & 034903+2431 &w&K5&4400 &    0.50 &   4.3 &   0.37$\pm$0.08 &    2.6$\pm$0.30 & B91&   (2.92,2.68)$\pm$0.22   & a,b \\
 &  &&K5&4100 &    0.50 &   4.14 &   0.33$\pm$0.02 &    1.73$\pm$0.30 & M92&   (2.40,2.14)$\pm$0.11   & a,b \\
&  & &K5&4391 &    0.32 &   4.45 &   0.31$\pm$0.02 &    2.3$\pm$0.35 & M94&   (2.62,2.30)$\pm$0.15   & a,b \\
&  & &K5&4391 &    0.32 &   4.45 &   0.35$\pm$0.02 &    2.3$\pm$0.35 & M94&   (2.85,2.56)$\pm$0.14   & a,b \\
&  & &K5&4350 &    0.44 &   4.30 &   0.37$\pm$0.08 &      & us&   (2.93,2.62)$\pm$0.20  & a,b \\
&  & &K5&4350 &    0.44 &   4.30 &   0.33$\pm$0.02 &      & us&   (2.70,2.38)$\pm$0.14  & a,b \\
&  & &K5&4350 &    0.44 &   4.30 &   0.31$\pm$0.02 &      & us&   (2.57,2.25)$\pm$0.14  & a,b \\
&  & &K5&4350 &    0.44 &   4.30 &   0.35$\pm$0.02 &      & us&   (2.81,2.50)$\pm$0.13  & a,b \\
 & & & & & & & & & & & \\
3 &035120+3154SW&w&G0&5900  &    0.9 &   4.5 &   0.29 &    4.6 & B91& 4.58,4.14 & a,b \\
 &&&G0&5943  &    0.77 &   4.63 &   0.13$\pm$0.01 &    3.10$\pm$0.35 & M94&(3.23,3.02)$\pm$0.10  & a,b \\
 &&&G0&6030  &    0.77 &   5.12 &   0.29 &    & us&4.62,4.26  & a,b \\
 &&&G0&6030  &    0.70 &   5.22 &   0.13$\pm$0.01 &    & us&(3.31,3.10)$\pm$0.10  & a,b \\
 & & & & & & & & & & & \\
4 &035120+3154NE&w&G5&5660  &    0.7 &   4.5 &   0.26 &    3.6 & B91& 4.14,3.62 & a,b \\
 &&&G5&5200  &    0.58 &   4.78 &   0.26$\pm$0.01 &    3.14$\pm$0.29 & M92&   (3.62,3.07)$\pm$0.09   & a,b \\
 &&&G5&5770  &    0.70 &   5.07 &   0.26 &    & us&   4.15,3.74   & a,b \\
 &&&G5&5770  &    0.58 &   5.17 &   0.26$\pm$0.01 &    & us&   (4.15,3.74)$\pm$0.08   & a,b \\
 & & & & & & & & & & & \\
5 &035135+2528NW&w&K3&4750&     0.2&    4.7 &   0.22 & 2.2 & B91& 2.54,2.20  & a,b \\
  &&&K3&4690&     0.22&    4.60 &   0.24$\pm$0.01 &    2.2$\pm$0.35 & M94&   (2.59,2.24)$\pm$0.08   & a,b \\
 & & &K3&4730&     0.22  &    4.60   &   0.22 && us& 2.51,2.17  & a,b \\
 & & &K3&4730&     0.22  &    4.60   &   0.24$\pm$0.01 && us& (2.65,2.30)$\pm$0.08  & a,b \\
 & & & & & & & & & & & \\
6 &035135+2528SE&w&K2&4950&     0.3&    4.6&   0.24 &    2.4 & B91& 3.02,2.60  & a,b \\
  &&&K1&4500&     0.30&    4.46&   0.23$\pm$0.01 &    1.87$\pm$0.29 & M92&   (2.21,1.91)$\pm$0.07   & a,b \\
& & &K2&4835&     0.33&    4.54&   0.29$\pm$0.1 &    2.7$\pm$0.35 & M94&   (3.18,2.80)$\pm$0.66   & a,b \\
& & &K0&5250&     0.34&    4.73&   0.24 &      & us& 3.44,2.98   & a,b \\
& & &K0&5250&     0.34&    4.73&   0.23$\pm$0.01 &      & us&   (3.38,2.92)$\pm$0.08   & a,b \\
& & &K0&5250&     0.34&    4.73&   0.29$\pm$0.1 &      & us&   (3.80,3.35)$\pm$0.71   & a,b \\
 & & & & & & & & & & & \\
7 &SAO 76411A &w&G1&5840 &    5.1 &   4.0 &  0.15& 3.0 & B91& 3.36,3.04  & a,b \\ 
 & &&G1&5945 &    5.1 &   3.9 &  0.15& & us& 3.44,3.14  & a,b \\ 
 & & & & & & & & & & & \\
8 &040012+2545N+S &w&K2&4839 &    0.16 &   4.81 &  0.36$\pm$0.04& 3.35$\pm$0.35& M94&(3.49,3.29)$\pm$0.20  & a,b \\ 
 & &&K2&4900 &    0.16 &   5.21 &  0.36$\pm$0.04& & us&(3.56,3.38)$\pm$0.22  & a,b \\ 
 & & & & & & & & & & & \\
9 &040047+2603W &w&M2&3524 &    0.20 &   3.90 &  $<$0.07& $<-$0.2& M94&  $<-$0.50     & c \\ 
& & &M2&3560 &    0.28 &   3.78 &  $<$0.07&     & us&  $<-$0.30     & c \\ 
& & & & & & & & & & & \\
10 &040047+2603E &w&M2&3524 &    0.20 &   3.90 &  $<$0.06& $<-$0.2& M94&  $<-$0.50     & c \\ 
& & &M2&3560 &    0.26 &   3.88 &  $<$0.06&     & us&   $<-$0.50    & c \\
 & & & & & & & & & & & \\
11 &SAO 76428 &w&F8&6000 &    3.0 &   4.1 &  0.18& 3.4 & B91& 3.79,3.38 & a,b \\ 
 & &&F8&6200 &    3.0 &   4.6 &  0.18& & us& 3.94,3.56  & a,b \\ 
& & & & & & & & & & & \\
12 &040142+2150SW&w&M3&3404 &    0.17 &   3.69 &  $<$0.25&  $<$0.50 & M94 &  $<$1.44    & c \\
& & &M3.5&3345 &  0.12 &   3.78	&  $<$0.25&     & us &    $<$1.34  & c \\ 
& & & & & & & & & & & \\
13 &040142+2150NE&w&M3&3404 &    0.16 &   3.72 &  $<$0.30&  $<$1.0  & M94 &   $<$1.78    & c \\ 
& & &M3&3415 &    $>$0.05 & $<$4.23 &  $<$0.30&   & us &   $<$1.74    & c \\ 
& & & & & & & & & & & \\
14 &040234+2143  &w&M2&3524 &    0.18 &   3.90 &  0.34$\pm$0.02   &   1.3$\pm$0.35 & M94 &    1.85$\pm$0.16      & c \\ 
&  & &M2&3560 &    0.13 &   4.06 &  0.34$\pm$0.02   &    & us &   1.84$\pm$0.17       & c \\ 
& & & & & & & & & & & \\
15 &LkCa 1      &w&M4&3288 &    0.66 &   3.11 &   0.56$\pm$0.03 &   --   & M94&   $>$3.5	     & c \\
&  & &M4&3270 &    0.38 &   3.34 &   0.56$\pm$0.03 &      & us&   $>$3.5	     & c \\
& & & & & & & & & & & \\
16 &V773 Tau~A &w&K2  &4750  &	   5.80 &   3.63 &   0.47 &    3.9 & B91&   4.23,3.69   & a,b \\
& & & K3 &4730  &    5.60 &   3.64 &   0.47 &	   & us&   4.19,3.65	& a,b \\
& & & & & & & & & & & \\
17 &LkCa 3   &w&M1  &3680  &	   1.70 &   3.10 &   0.57 &    1.8 & B91&   $>$3.5       & c \\
& & &M1  &3657  &    0.98 &   3.33 &   0.55$\pm$0.03 &	 3.1$\pm$0.35 & M94&   $>$3.5      & c \\
& & &M1  &3705  &    1.66 &   3.05 &   0.57 &	   & us&   $>$3.5	& c \\
& & &M1  &3705  &    1.66 &   3.05 &   0.55$\pm$0.03 &	  & us&   $>$3.5     & c \\
& & & & & & & & & & & \\
18 &LkCa 4   &w&K7  &4000  &	   0.90 &   3.73 &   0.51 &    2.6 & B91&   3.11,2.80   & a,b \\
& & & K7 &4130  &    0.89 &   3.79 &   0.71$\pm$0.04 &	3.3$\pm$0.35  & M94&   (3.72,3.31)$\pm$0.07	& a,b \\
& & & K7 &4130  &    0.89 &   3.79 &   0.51$\pm$0.03 &	3.3$\pm$0.35  & M94&   (3.29,2.90)$\pm$0.09	& a,b \\
& & & K7 &4060  &    0.85 &   3.78 &   0.51 &	   & us&   3.19,2.85 & a,b \\
& & & K7 &4060  &    0.85 &   3.78 &   0.51$\pm$0.03 &	   & us&   (3.19,2.85)$\pm$0.10	& a,b \\
& & & K7 &4060  &    0.85 &   3.78 &   0.71$\pm$0.04 &	   & us&   (3.61,3.25)$\pm$0.06	& a,b \\
& & & & & & & & & & & \\
19 &LkCa 5      &w&M2&3522 &    0.38 &   3.62 &   0.55$\pm$0.03&	2.5$\pm$0.35  & M94&    $>$3.5	& c \\
& & &M2&3560 &    0.37 &   3.65 &   0.55$\pm$0.03 &	   & us&    $>$3.5	& c \\
& & & & & & & & & & & \\
20 &041529+1652  &w&K5&4400 &    0.17 &   4.61 &  0.19$\pm$0.01   &   1.5$\pm$0.35 & M94 &    (1.82,1.52)$\pm$0.06 & a,b \\
& & &K5&4350 &    0.14 &   4.67 &  0.19$\pm$0.01   &     & us &   (1.77,1.46)$\pm$0.07  & a,b \\
& & & & & & & & & & & \\
21 &V410 Tau~ABC &w& K3 & 4750 &   1.90 &   3.69  &   0.42$\pm$0.3 &	3.55$\pm$0.1 & B91&   (3.98,3.48)$\pm$0.15	& a,b	 \\
& & & K7 &3900  &   2.40 &   3.24  &   0.57$\pm$0.05 &	 2.76$\pm$0.20 & M92&   (3.32,2.86)$\pm$0.12      & a,b \\
& & & K7 & 4060 &   1.50 &   3.52  &   0.42$\pm$0.3 &	  & us&   (2.97,2.55)$\pm$0.13	& a,b \\
& & & K7 &4060  &   1.50 &   3.52  &   0.57$\pm$0.05 &	   & us&   (3.46,2.99)$\pm$0.12	& a,b \\
& & & & & & & & & & & \\
22 &CZ Tau~AB   &w&M1&3590 &    0.24 &   3.83 &   0.46$\pm$0.02 &	2.1$\pm$0.35  & M94&   3.05$\pm$0.15      & c \\
& & &M3&3415 &    0.27 &   3.70 &   0.46$\pm$0.02 &	   &us&   3.00$\pm$0.15      & c \\
& & & & & & & & & & & \\
23 &Hubble 4    &w&K7&4133 &    0.95 &   3.75 &   0.61$\pm$0.03 &	3.3$\pm$0.35  & M94&   (3.55,3.15)$\pm$0.07	& a,b \\
&  & &K7&4060 &    2.70 &   3.26 &   0.61$\pm$0.03 &	   & us&   (3.64,3.08)$\pm$0.07	& a,b \\  
& & & & & & & & & & & \\
24 &Anon 1      &w&M0&3830 &    0.77 &   3.59 &   0.48$\pm$0.02 &	2.6$\pm$0.35  & M4&   (2.87,2.57)$\pm$0.07      & a,b \\
& & &M0&3850 &    2.60 &   3.07 &   0.48$\pm$0.02 &	   & us&   (3.07,2.61)$\pm$0.07     & a,b \\
& & & & & & & & & & & \\
25 &041559+1716 &w&K7&4000 &    0.5 &   4.0 &   0.53 &	2.6  & B91& 3.08,2.85 & a,b \\
 & &&K7&3900 &    0.48 &   3.98 &   0.60$\pm$0.08 &	2.70$\pm$0.27  & M92&   (3.15,2.92)$\pm$0.17      & a,b \\
 & &&K7&4142 &    0.4 &   4.14 &   0.46$\pm$0.05 &	2.70$\pm$0.35  & M94& (3.04,2.76)$\pm$0.25  & a,b \\
& & &K7&4060 &    0.41 &   4.12 &   0.53 &   & us& 3.13,2.90  	& a,b \\
& & &K7&4060 &    0.41 &   4.12 &   0.60$\pm$0.08 & 	   & us&   (3.29,3.06)$\pm$0.16	& a,b \\
& & &K7&4060 &    0.41 &   4.12 &   0.46$\pm$0.05 &	   & us& (2.92,2.69)$\pm$0.17  & a,b \\
& & & & & & & & & & & \\
26 &BP Tau   &c& K7 & 4000 &   0.90 &   3.73  &   0.69$\pm$0.03 &	3.15$\pm$0.1 & B91&   (3.49,3.17)$\pm$0.06	& a,b	 \\
& & & K7 &3900  &   1.62 &   3.42  &   0.62$\pm$0.04 &	 2.94$\pm$0.23 & M92&   (3.37,2.79)$\pm$0.08      & a,b \\
& & & K7 & 4060 &   0.95 &   3.73  &   0.69$\pm$0.03 &	   & us&   (3.59,3.22)$\pm$0.05	& a,b \\
& & & K7 &4060  &   0.95 &   3.73  &   0.62$\pm$0.04 &	   & us&   (3.47,3.10)$\pm$0.07	& a,b \\
& & & & & & & & & & & \\
27 &V819 Tau~AB  &w&K7&4136 &    0.80 &   3.84 &   0.62$\pm$0.03 &	3.3$\pm$0.35  & M94&   (3.54,3.17)$\pm$0.07	& a,b \\
& & &K7&4060 &    0.91 &   3.75 &   0.62$\pm$0.03 &	   & us&   (3.46,3.10)$\pm$0.07	& a,b \\  
& & & & & & & & & & & \\
28 &LkCa 7~A  &w&M0  &4000  &	   0.80 &   3.78 &   0.60 &    2.9 & B91&   3.31,3.02   & a,b \\
& & &M0  &3600  &    0.98 &   3.51 &   0.63$\pm$0.4 &	 2.29$\pm$0.22 & M92&   $>$3.5      & c \\
& & & K7 &4133  &    0.60 &   3.96 &   0.63$\pm$0.03 &	3.3$\pm$0.35  & M94&   (3.52,3.18)$\pm$0.07	& a,b \\
& & & K7 &4133  &    0.60 &   3.96 &   0.52$\pm$0.03 &	3.3$\pm$0.35  & M94&   (3.27,2.93)$\pm$0.09	& a,b \\
& & & K7 &4060  &    1.20 &   3.63 &   0.60 &	   & us&   3.47,3.06	& a,b \\
& & & K7 &4060  &    1.20 &   3.63 &   0.63$\pm$0.40 &	   & us&   (3.52,3.12)$\pm$0.50	& a,b \\
& & & K7 &4060  &    1.20 &   3.63 &   0.63$\pm$0.03 &	   & us&   (3.53,3.12)$\pm$0.08	& a,b \\
& & & K7 &4060  &    1.20 &   3.63 &   0.52$\pm$0.03 &	   & us&   (3.27,2.87)$\pm$0.09	& a,b \\
& & & & & & & & & & & \\
29 &DE Tau   &c&M2  &3500  &   0.60 &   3.41  &   0.99$\pm$0.13 &	 2.3$\pm$0.2 & B91&   $>$3.5     & c \\
& & &M2  &3560  &   0.81 &   3.31  &   0.99$\pm$0.13 &	   & us&   $>$3.5      & c \\
& & & & & & & & & & & \\
30 &RY Tau   &c&K0  &5250  &   3.00 &   4.10  &   0.27$\pm$0.04 &	 3.3$\pm$0.2 & B91&   (3.76,3.21)$\pm$0.31	& a,b \\
& & &G1  &5945  &   7.60 &   3.98  &   0.27$\pm$0.04 &	   & us&   (4.61,3.99)$\pm$0.31	& a,b \\
& & & & & & & & & & & \\
31 &T Tau    &c&K0  &5250  &   10.80&    3.54 &   0.34$\pm$0.05 &	 3.8$\pm$0.25 & B91&   (4.33,3.73)$\pm$0.32	& a,b \\
& & &K0  &5250  &   19.90&    3.36 &   0.34$\pm$0.05 &	   & us&   (4.35,3.73)$\pm$0.32	& a,b \\
& & & & & & & & & & & \\
32 &LkCa 21     &w&M3&3400 &    0.60 &   3.29 &   0.72$\pm$0.04 &	2.9$\pm$0.35  & M94&   $>$3.5     & c \\
& & &M3&3415 &    0.62 &   3.28 &   0.72$\pm$0.04 &	   & us&   $>$3.5      & c \\
& & & & & & & & & & & \\
33 &IP Tau      &w&M0&3832 &    0.50 &   3.79 &   0.50$\pm$0.03 &	2.6$\pm$0.35  & M94&   (2.88,2.63)$\pm$0.09      & a,b \\
& & &M0&3850 &    0.43 &   3.86 &   0.50$\pm$0.03 &	   & us&   (2.88,2.65)$\pm$0.10      & a,b \\
& & & & & & & & & & & \\
34 &DF Tau~B  &c&M2  &3500  &	   1.90&    2.82 &   0.79$\pm$0.10 &    2.0$\pm$0.2 & B91&   $>$3.5       & c \\
& & &M3  &3415  &    1.50&    2.88 &   0.79$\pm$0.10 &	   & us&   $>$3.5      & c \\
& & & & & & & & & & & \\
35 &DG Tau   &c&M0  &3920  &	   1.50&    3.73 &   0.51$\pm$0.16 &    2.5$\pm$0.5 & B91&   (3.02,2.73)$\pm$0.36        & a,b \\
& & &M0  &3850  &    1.70&    3.65	 &   0.51$\pm$0.16   &     & us&   (2.96,2.67)$\pm$0.33      & a  \\
& & & & & & & & & & & \\
36 &042417+1744&w&K1&5250  &    1.6 &   4.2 &   0.28$\pm$0.09 &	 3.40$\pm$0.45 & B91& (3.81,3.28)$\pm$0.60  & a,b \\
 &&&K1&4600  &    1.66 &   4.12 &   0.27$\pm$0.05 &	 2.32$\pm$0.29 & M92&   (2.62,2.32)$\pm$0.36	& a,b \\
 &&&K1&4913  &    0.89 &   4.29 &   0.28$\pm$0.09 &	 2.80$\pm$0.35 & M94& (3.24,2.84)$\pm$0.55   	& a,b \\
& & &K1&5080  &    1.40 &   4.19 &   0.28$\pm$0.09 &	   & us&   (3.54,3.06)$\pm$0.60	& a,b \\
& & &K1&5080  &    1.40 &   4.19 &   0.27$\pm$0.05 &	   & us&   (3.47,2.99)$\pm$0.36	& a,b \\
& & &K1&5080  &    1.40 &   4.19 &   0.28$\pm$0.09 &	   & us&   (3.54,3.06)$\pm$0.60	& a,b \\
& & & & & & & & & & & \\
37 &DI Tau~AB    &w&M0&3826 &    0.73 &   3.61 &   0.69$\pm$0.03 &	3.3$\pm$0.35  & M94&   (3.32,3.02)$\pm$0.05      & a,b \\
& & &M0&3850 &    0.99 &   3.49 &   0.69$\pm$0.03 &	   & us&   (3.40,3.05)$\pm$0.06      & a,b \\
& & & & & & & & & & & \\
38 &UX Tau A &w&K0  &5250  &	   1.00&    4.41 &   0.42$\pm$0.05 &    4.2$\pm$0.2 & B91&   (4.51,4.19)$\pm$0.26   & a,b \\
& & &K2  &4740  &    1.30&    4.12 &   0.43$\pm$0.02 &	 3.6$\pm$0.35 & M94&   (3.88,3.51)$\pm$0.10	& a,b \\
& & &K5  &4350  &    0.90&   4.13 &   0.42$\pm$0.05 &	  & us&	 (3.20,2.85)$\pm$0.26  & a,b \\
& & &K5  &4350  &    0.90&    4.13 &   0.43$\pm$0.02 &	   & us&   (3.24,2.89)$\pm$0.10	& a,b \\
& & & & & & & & & & & \\
39 &UX Tau B &w	&M1&3650 &    0.50&    3.57 &	0.60$\pm$0.03 &   3.1$\pm$0.35  & M94&   $>$3.5 & c \\
& & &M2&3560 &    0.19&    3.94 &	0.60$\pm$0.03 &      & us&   $>$3.5	      & c \\   
& & & & & & & & & & & \\
40 &DK Tau~AB &c&M0  &3920  &    1.40&    3.49 &   0.65$\pm$0.06 &	 2.9$\pm$0.2 & B91&   (3.43,3.04)$\pm$0.12      & a,b \\
& & & K7 &4060  &    1.32&    3.58 &   0.65$\pm$0.06 &	   & us&   (3.59,3.15)$\pm$0.12   & a,b \\
& & & & & & & & & & & \\
41 &V927 Tau~AB &w&M5.5&3101 &    0.36 &   3.75 &   0.51$\pm$0.03 &	--   & M94&   2.87$\pm$0.26     & c \\ 
& & &M5.5&3115 &    0.33 &   3.21 &   0.51$\pm$0.03 &	  & us&   2.90$\pm$0.26      & c \\ 
& & & & & & & & & & & \\
42 &042835+1700 &w&K5&4400 &    0.40 &   4.3 &   0.15 &	1.3  & B91& 1.45,1.31   & a,b \\
 & &&K5&4100 &    0.43 &   4.16 &   0.19$\pm$0.04 &	0.97$\pm$0.35  & M92&   (1.41,1.19)$\pm$0.24	& a,b \\
& & &K5&4400 &    0.20 &   4.60 &   0.11$\pm$0.01 &	1.15$\pm$0.35 & M94&   (1.20,1.08)$\pm$0.07	& a,b \\
& & &K5&4350 &    0.32 &   4.39 &   0.15 &	   & us& 1.39,1.24  	& a,b \\
& & &K5&4350 &    0.32 &   4.39 &   0.19$\pm$0.04 &	   & us&   (1.72,1.46)$\pm$0.23	& a,b \\
& & &K5&4350 &    0.32 &   4.39 &   0.11$\pm$0.01 &	   & us&   (1.11,1.01)$\pm$0.08	& a,b \\
& & & & & & & & & & & \\
43 &HL Tau   &c& K7 &4000  &	   1.40&    3.66 &   0.48$\pm$0.12 &    2.4$\pm$0.5 & B91&   (3.12,2.71)$\pm$0.31   & a,b \\
& & &K5  &4350  &    1.53&    3.67 &   0.48$\pm$0.12 &	  & us&   (3.59,3.07)$\pm$0.32	& a,b \\
& & & & & & & & & & & \\
44 &V710 Tau B  &w&M3&3402 &    0.30 &   3.61 &   0.58$\pm$0.03&	2.4$\pm$0.35  & M94&   $>$3.5     & c \\ 
& & &M2&3560 &    0.63 &   3.37 &   0.58$\pm$0.03 &	   & us&   $>$3.5      & c \\   
& & & & & & & & & & & \\
45 &042916+1751 &w&K7&4138 &    0.62 &   3.85 &   0.49$\pm$0.05&	3.15$\pm$0.35  & M94& (3.22,2.85)$\pm$0.16  & a,b \\ 
 & &&K7&4060 &    0.62 &   4.25 &   0.49$\pm$0.05&	& us& (2.96,2.79)$\pm$0.15  & a,b \\ 
& & & & & & & & & & & \\
46 &V827 Tau &w&K7  &3960  &	   0.90 &   3.71 &   0.57 &    2.8 & B91&   3.22,2.91        & a,b \\
& & & K7 &4130  &    1.11 &   3.69 &   0.57$\pm$0.03 &	3.3$\pm$0.35  & M94&   (3.49,3.06)$\pm$0.08	& a,b \\
& & & K7 &4060  &    1.10 &   3.67 &   0.57 &	   & us&   3.38,2.99	& a,b \\
& & & K7 &4060  &    1.10 &   3.67 &   0.57$\pm$0.03 &	   & us&   (3.38,2.99)$\pm$0.08	& a,b \\
& & & & & & & & & & & \\
47 &V928 Tau~AB &w&M0.5&3745 &    1.30 &   3.29 &   0.64$\pm$0.03 &	3.1$\pm$0.35  & M94&   $>$3.5      & c \\ 
& & &M0.5&3778 &    1.40 &   3.27 &   0.64$\pm$0.03 &	   & us&   $>$3.5      & c \\ 
& & & & & & & & & & & \\
48 &GG Tau~A  &c& K7 &4000  &	   1.20&    3.50 &   0.72$\pm$0.03 &    3.2$\pm$0.1 & B91&   (3.65,3.22)$\pm$0.06   & a,b \\
& & & K7 &4060  &    1.50&    3.50 &   0.72$\pm$0.03 &	   & us&   (3.76,3.26)$\pm$0.05   & a,b \\
& & & & & & & & & & & \\
49 &UZ Tau W &c&M2  &3500  &	   1.10&    3.11 &   0.74$\pm$0.18 &    1.8$\pm$0.3 & B91&   $>$3.5        & c \\
& & &M3  &3470  &    0.52&    3.42 &   0.74$\pm$0.18 &	   & us&   $>$3.5      & c \\
& & & & & & & & & & & \\
50 &UZ Tau E &c&M0  &3920  &	   1.00&    3.44 &   0.72$\pm$0.15 &    3.0$\pm$0.3 & B91&   (3.56,3.16)$\pm$0.31        & a,b \\
& & &M1  &3705  &    0.40&    3.74 &   0.72$\pm$0.15 &	   & us&   $>$3.5     & c \\
& & & & & & & & & & & \\
51 &042950+1757 &w&K7&4000 &    0.4 &   4.1 &   0.54 &	2.6  & B91& 3.07,2.87  & a,b \\
 & &&K7&3900 &    0.44 &   4.12 &   0.60$\pm$0.06 &	2.72$\pm$0.24  & M92&   (3.11,2.92)$\pm$0.13      & a,b \\
& & &K7&4060 &    0.37 &   4.16 &   0.54 &	   & us& 3.14,2.92  & a,b \\
& & &K7&4060 &    0.37 &   4.16 &   0.60$\pm$0.06 &	   & us&   (3.27,3.06)$\pm$0.14	& a,b \\
& & & & & & & & & & & \\
52 &GH Tau~AB      &w&M2&3520 &    0.89 &   3.26 &   0.66$\pm$0.03 &	2.9$\pm$0.35  & M94&   $>$3.5      & c \\
& & &M1.5&3632 &    0.81 &   3.35 &   0.66$\pm$0.03 &	   & us&   $>$3.5      & c \\
& & & & & & & & & & & \\
53 &V830 Tau &w& K7 &4000  &	   0.90&    3.73 &   0.70$\pm$0.06 &    3.0$\pm$0.2 & B91&   (3.51,3.19)$\pm$0.11   & a,b \\
& & & K7 &4134  &    0.89&    3.80 &   0.65$\pm$0.03 &	 3.5$\pm$0.35 & M94&   (3.61,3.22)$\pm$0.06	& a,b \\
& & & K7 &4060  &    0.78&    3.82 &   0.70$\pm$0.06 &	   & us&   (3.57,3.23)$\pm$0.11	& a,b \\
& & & K7 &4060  &    0.78&    3.82 &   0.65$\pm$0.03 &	   & us&   (3.49,3.15)$\pm$0.06	& a,b \\
& & & & & & & & & & & \\
54 &IS Tau      &w&K7&4130 &    1.10 &   3.71 &   0.62$\pm$0.03 &	3.2$\pm$0.35  & M94&   (3.59,3.16)$\pm$0.07	& a,b \\
& & &K7&4060 &    0.66 &   3.90 &   0.62$\pm$0.03 &	   & us&   (3.40,3.10)$\pm$0.07	& a,b \\
& & & & & & & & & & & \\
55 &DL Tau   &c&M0  &3920  &	   0.90&    3.51 &   0.80$\pm$0.12 &    3.2$\pm$0.3 & B91&   (3.66,3.28)$\pm$0.15        & a,b \\
& & & K7 &4060  &    1.16&    3.49 &   0.80$\pm$0.12 &	   & us&   (3.88,3.38)$\pm$0.16	& a,b \\
& & & & & & & & & & & \\
56 &CI Tau   &c&K5  &4400  &	   0.60&    4.07 &   0.57$\pm$0.06 &    3.5$\pm$0.2 & B91&   (3.77,3.38)$\pm$0.16   & a,b \\
& & & K7 &4060  &    0.87&    3.77 &   0.57$\pm$0.06 &	   & us&   (3.34,2.99)$\pm$0.14	& a,b \\
& & & & & & & & & & & \\
57 &AA Tau   &c&M0  &3920  &	   0.70&    3.81 &   0.64$\pm$0.03 &    2.9$\pm$0.1 & B91&   (3.29,3.03)$\pm$0.06        & a,b \\
& & & K7 &4060  &    0.80&    3.81 &   0.64$\pm$0.03 &	   & us&   (3.47,3.14)$\pm$0.06	& a,b \\
& & & & & & & & & & & \\
58 &DN Tau   &c&M0  &3920  &	   1.00&    3.52 &   0.65$\pm$0.05 &    2.9$\pm$0.15 & B91&   (3.42,3.04)$\pm$0.10        & a,b \\
& & &M0  &3850  &    0.65&    3.63 &   0.65$\pm$0.05 &	   & us&   (3.27,2.97)$\pm$0.11      & a,b \\
& & & & & & & & & & & \\
59 &043124+1824&w&G8  & 5450 &    0.5 &   4.6 &	0.31 &    3.8& B91 & 4.20,3.77   & a,b \\
 &&&G8  & 5445 &    0.47 &   4.65 &	0.23$\pm$0.01 &    3.10$\pm$0.35& M94 & (3.62,3.15)$\pm$0.08   & a,b \\
 &&&G8  & 5520 &    0.5 &   5.0 &	0.31 &    & us & 4.18,3.85   & a,b \\
 &&&G8  & 5520 &    0.47 &   5.1 &	0.23$\pm$0.01 &    & us & (3.65,3.23)$\pm$0.07   & a,b \\
& & & & & & & & & & & \\
60 &043220+1815&w&F8  & 6000 &    1.5 &   4.4 &	0.17 &    3.4& B91 & 3.68,3.32   & a,b \\
 &&&F8  & 6200 &    1.5 &   5.0 &	0.17 &    & us & 3.83,3.50   & a,b \\
& & & & & & & & & & & \\
61 &043230+1746&w&--  & 3500 &    0.40 &   4.22 &	0.57$\pm$0.03 &    1.4& B91 &    $>$3.5	& c \\
& & &M2&3524 &    0.22 &   3.84 &  0.57$\pm$0.03   &   2.6$\pm$0.35 & M94&    3.31$\pm$0.25	 & c \\
& & &M2&3560 &    0.43 &   3.57 &	0.57$\pm$0.03 &    & us &  3.34$\pm$0.25 	& c \\\
& & & & & & & & & & & \\
62 &LkCa 14     &w&K7&4134 &    0.88 &   3.67 &   0.60$\pm$0.03 &	3.3$\pm$0.35  & M94&   (3.57,3.13)$\pm$0.06	& a,b \\
& & &M0&3850 &    0.65 &   3.68 &   0.60$\pm$0.03 &	   & us&   (3.16,2.87)$\pm$0.07      & a,b \\
& & & & & & & & & & & \\
63 &HV Tau~AB   &w&M2&3521 &    0.65 &   3.37 &   0.64$\pm$0.03 &	2.8$\pm$0.35  & M94&   $>$3.5      & c \\
& & &M1& 3705 &     0.65 &   3.37 &   0.64$\pm$0.03 &	   & us&   $>$3.5      & c \\
& & & & & & & & & & & \\
64 &VY Tau~AB   &w&M0  &3920  &	   0.40 &   3.92 &   0.52 &    2.4 & B91&   3.00,2.76        & a \\
& & &M0  &3850  &    0.47 &   3.82 &   0.52 &	   & us&   2.95,2.70      & a \\
& & & & & & & & & & & \\
65 &LkCa 15     &w&K5&4385 &    0.72 &   4.01 &   0.47$\pm$0.02 &	3.1$\pm$0.35  & M94&   (3.46,3.09)$\pm$0.07	& a,b \\
& & &K5&4350 &    0.74 &   4.13 &   0.47$\pm$0.02 &	   & us&   (3.38,3.04)$\pm$0.08	& a,b \\
& & & & & & & & & & & \\
66 &IW Tau      &w&K7&4130 &    1.10 &   3.70 &   0.44$\pm$0.02&	2.9$\pm$0.35  & M94&   (3.09,2.68)$\pm$0.08	& a,b \\
& & &K7&4060 &    0.87 &   3.77 &   0.44$\pm$0.02 &	   & us&   (2.96,2.63)$\pm$0.07	& a,b \\
& & & & & & & & & & & \\
67 &DR Tau   &c&M0  &3920  &	   1.70&    3.30 &   0.46$\pm$0.06 &    2.1$\pm$0.25 & B91&   (2.99,2.59)$\pm$0.19        & a,b \\
& & & K7 &4060  &    0.96&    3.88 &   0.46$\pm$0.06 &	   & us&   (3.00,2.69)$\pm$0.20	& a,b \\
& & & & & & & & & & & \\
68 &DS Tau   &c&K5  &4400  &    0.80&    4.10 &   0.66$\pm$0.09 &	 3.7$\pm$0.25 & B91&   (3.94,3.59)$\pm$0.20	& a,b \\
& & &K5  &4350  &    0.65&    4.17 &   0.66$\pm$0.09 &	   & us&   (3.82,3.51)$\pm$0.20	& a,b \\
& & & & & & & & & & & \\
69 &UY Aur~AB   &c&M0  &3920  &	   1.40&    3.49 &   0.87$\pm$0.14 &    3.2$\pm$0.2 & B91&   (3.76,3.37)$\pm$0.21        & a,b \\
& & & K7 &4060  &    1.20&    3.62 &   0.87$\pm$0.14 &	   & us&   (3.90,3.47)$\pm$0.20	& a,b \\
& & & & & & & & & & & \\
70 &GM Aur   &c& K7 &4000  &	   0.90&    3.85 &   0.50$\pm$0.05 &    2.5$\pm$0.2 & B91&   (3.04,2.77)$\pm$0.15   & a,b \\
& & & K7 &3800  &    0.93 &   3.74 &   0.59$\pm$0.05 &	 2.56$\pm$0.24 & M92&   (3.07,2.81)$\pm$0.11      & a,b \\
& & & K7 &4060  &    0.67&    3.75 &   0.50$\pm$0.05 &	   & us&   (3.17,2.82)$\pm$0.14	& a,b \\
& & & K7 &4060  &    0.67 &   4.00 &   0.59$\pm$0.05 &	   & us&   (3.31,3.04)$\pm$0.12	& a,b \\
& & & & & & & & & & & \\
71 &LkCa 19     &w&K5&4348 &    1.55 &   3.90 &   0.48$\pm$0.02 &	3.1$\pm$0.35  & M94&   (3.51 3.07)$\pm$0.07	& a,b \\
& & &K5&4348 &    1.55 &   3.90 &   0.43$\pm$0.02 &	3.1$\pm$0.35  & M94&   (3.32 2.89)$\pm$0.10	& a,b \\
& & &K0&5250 &    1.70 &   4.19 &   0.48$\pm$0.02 &	   & us&   (4.79 4.48)$\pm$0.09	& a,b \\
& & &K0&5250 &    1.70 &   4.19 &   0.43$\pm$0.02 &	   & us&   (4.62 4.23)$\pm$0.10	& a,b \\
& & & & & & & & & & & \\
72 &045226+3013 &w&K5&4950 &    1.5 &   4.1 &   0.44$\pm$0.05 &	3.8$\pm$0.2  & B91& (4.25,3.86)$\pm$0.24  & a,b \\
 & &&K5&4100 &    1.58 &   3.60 &   0.47$\pm$0.04 &	2.55$\pm$0.34  & M92&   (3.20,2.76)$\pm$0.15	& a,b \\
 & &&K5&4350 &    1.5 &   4.4 &   0.44$\pm$0.05 &	& us& (3.18,2.93)$\pm$0.20  & a,b \\
 & &&K5&4350 &    1.58 &   4.3 &   0.47$\pm$0.04 &	& us& 2.99$\pm$0.19  & a,b \\
 & & & & & & & & & & & \\
73 &SU Aur   &c&G2  &5770  &	  12.10 &   3.62 &   0.24$\pm$0.02 &    3.55$\pm$0.1& B91&   (4.20,3.58)$\pm$0.16   & a,b \\
& & &G2  &5860  &    9.90 &   3.75 &   0.24$\pm$0.02 &	   & us&   (4.29,3.68)$\pm$0.15	& a,b \\
& & & & & & & & & & & \\
74 &045251+3016 &w&K7&4000 &    0.8 &   3.9 &   0.58 &	2.8 & B91& 3.16,2.96  & a,b \\
& &&K7&4130 &    1.0 &   3.63 &   0.52$\pm$0.03 &	3.15$\pm$0.35 &M94& (3.38,2.93)$\pm$0.08  & a,b \\
& &&K7&4060 &    0.8 &   4.2 &   0.58 &	&us& 3.17,3.02  & a,b \\
& &&K7&4060 &    1.0 &   4.1 &   0.52$\pm$0.03 &	&us& (3.04,2.87)$\pm$0.07  & a,b \\
& & & & & & & & & & & \\
75 &V836 Tau &c&K7  &3960  &	   0.50 &   3.93 &   0.57 &    2.7 & B91&   3.16,2.91        & a,b \\
& & & K7 &4139  &    0.60 &   3.93 &   0.57$\pm$0.02 &	3.2$\pm$0.35  & M94&   (3.41,3.06)$\pm$0.06	& a,b \\
& & & K7 &4060  &    1.21 &   3.59 &   0.57$\pm$0.02 &	   & us&   (3.42,2.99)$\pm$0.06	& a,b \\
& & & K7 &4060  &    1.21 &   3.60 &   0.57$\pm$0.02 &	   & us&   (3.41,2.99)$\pm$0.06	& a,b \\
& & & & & & & & & & & \\
76 &RW Aur~A   &c&K5  &4400  &	   2.20 &   3.87 &   0.67$\pm$0.07 &    3.9$\pm$0.2 & B91&   (4.07,3.62)$\pm$0.15   & a,b \\
& & &K3  &4730  &    1.70 &   4.11 &   0.67$\pm$0.07 &	   & us&   (4.52,4.13)$\pm$0.10	& a,b \\
\hline\hline
\multicolumn{12}{l}{Notes:		     	        		 
n(Li)$_{\rm our}$ -- our derivation of Li abundance and error, using the method listed in the next column}\\
\multicolumn{12}{l}{Method -- a: MOOG; b: COGs by Soderblom  et al.  (1993); c: our COGs (resolution depending on the literature data).}
\end{longtable}	
\begin{figure*}
\psfig{figure=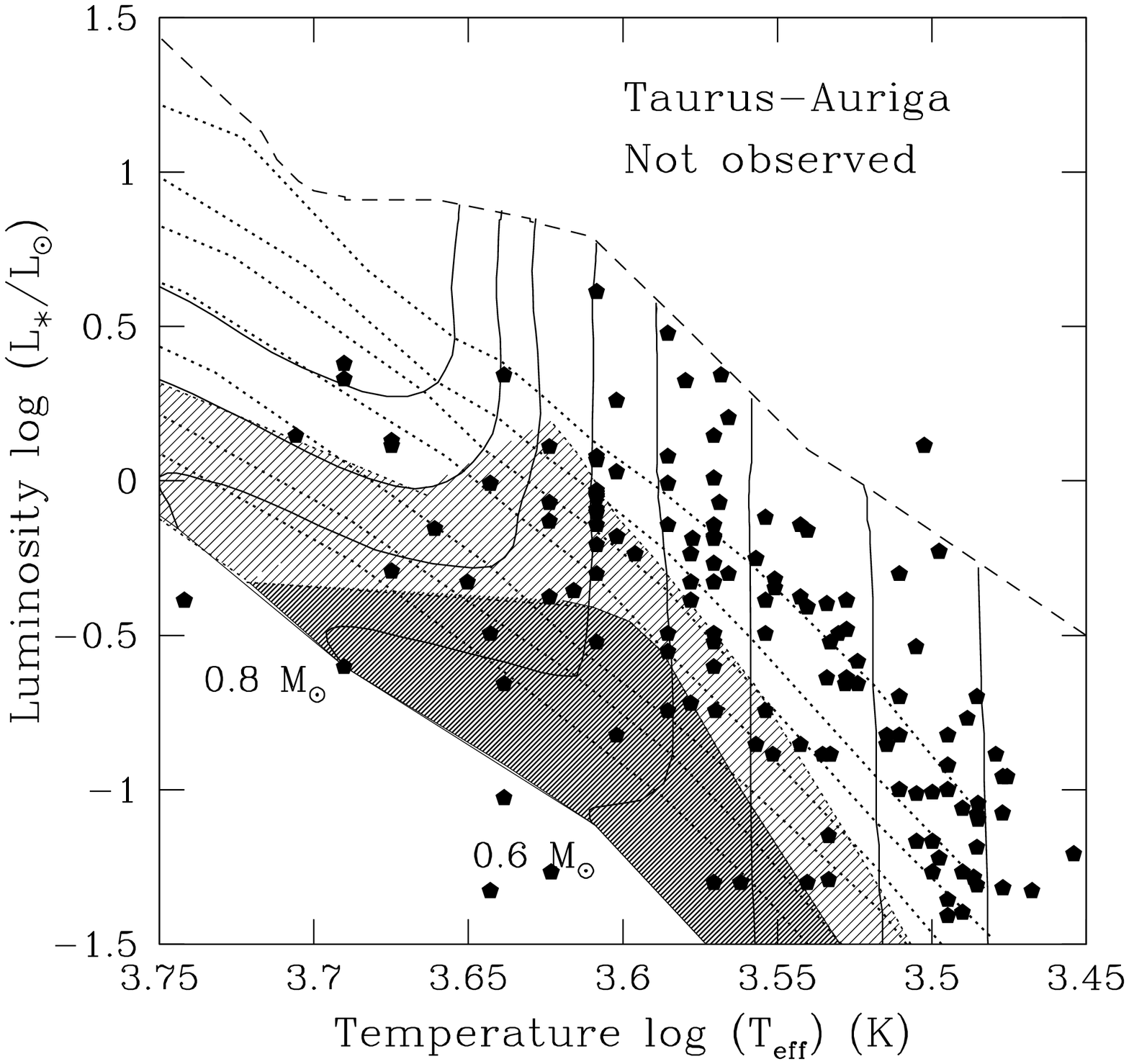, width=12cm, angle=0}
\caption{The distribution in the HR diagram of the known population of low-mass
stars of Tau-Aur without measurements of the Li abundance. The sample is based
on the compilations by Palla \& Stahler (2000) and G\"udel et al. (2007).}
\label{Li_future}
\end{figure*}

\begin{thebibliography}{}
\bibitem[1998]{baraffe98}Baraffe, I., Chabrier, G., Allard, F., \& Hauschildt, 
P.H. 1998, A\&A, 337, 403
\bibitem[2002]{baraffe02}Baraffe, I., Chabrier, G., Allard, F., \& Hauschildt, 
P.H. 2002, A\&A, 382, 563
\bibitem[2004]{byn} Barrado y Navascu\'es, D., Stauffer, J. R.,
Jayawardhana, R. 2004, ApJ, 614, 386
\bibitem[1990]{BB90}Basri, G., \& Batalha C. 1990, ApJ, 363, 654
\bibitem[1991]{basri}Basri, G., Mart\'{\i}n, E.L., \& Bertout, C. 1991, 
A\&A, 252, 625 (B91)
\bibitem[2006]{bg06}Bertout, C. \& Genova, F. 2006, A\&A, 460, 499
\bibitem[2007]{bertout07}Bertout, C., Siess, L., \& Cabrit, S. 2007, A\&A, 473, L21
\bibitem[1997]{bildsten}Bildsten, L., Brown, E.F., Matzner, C.D., Ushomirsky,
G. 1997, ApJ, 482, 442
\bibitem[2002]{briceno97}Brice\~no, C., Hartmann, L.W., Stauffer, J.H., 
et al. 1997, AJ, 113, 740
\bibitem[2002]{briceno02}Brice\~no, C., Luhman, K.L., Hartmann, L., Stauffer,
J.R., Kirkpatrick, J.D. 2002, ApJ, 580, 317
\bibitem[1979]{CK79}Cohen, M., \& Kuhi, L. 1979, ApJS, 41, 743
\bibitem[1994]{dantona94}D'Antona, F., \& Mazzitelli, I. 1994, ApJS, 90, 467
\bibitem[2003]{dantona03}D'Antona, F., \& Montalb\`an, J. 2003, A\&A, 412, 213
\bibitem[1987]{dejager}de Jager, C., \& Nieuwenhuijzen, H. 1987, A\&A, 177,
217
\bibitem[2008]{dorazi}D'Orazi, V., Randich, S., Flaccomio, E., et al. 2008,
submitted to A\&A
\bibitem[2005]{doucourant}Doucourant, C., Teixeira, R., P\'eri\'e, J.P., et al.
2005, A\&A, 438, 769
\bibitem[2007]{XEST}G\"udel, M., Briggs, K.R., Arzner, K., et al. 2007, A\&A, 468, 353
\bibitem[1986]{hartmann}Hartmann, L.W., Hewett, R., Stahler, S., Mathieu, R.D.
1986, ApJ, 309, 275
\bibitem[2006]{jeff06} Jeffries, R. D.  2006, in Chemical Abundances and Mixing
in the Milky Way and its Satellites, S. Randich \& L. Pasquini (eds), ESO 
Astrophysics Symposia, Springer-Verlag, p.~163
\bibitem[2005]{jo05} Jeffries, R. D. \& Oliveira, J. M. 2005, MNRAS, 358, 13
\bibitem[1995]{KH95}Kenyon, A.J., \& Hartmann, L. 1995, ApJS, 101, 177
\bibitem[1995]{kuru}Kurucz R.L. 1995, ApJ, 452, 102
\bibitem[1999]{luhman99}Luhman, K.L. 1999, ApJ, 525, 466
\bibitem[2003]{luhman03}Luhman, K.L., Brice\~no, C., Stauffer, J.R., 
et al. 2003, ApJ, 590, 348
\bibitem[1992]{magazzu}Magazz\`u, A., Rebolo, R., \& Pavlenko, Ya.V. 1992, 
ApJ, 392, 159 (M92)
\bibitem[2008]{manzi08}Manzi, S., Randich, S., de Wit, W.-J., Palla, F.
2008, A\&A, 479, 141
\bibitem[1998]{martin98}Mart\'{\i}n, E.L. 1998, AJ, 115, 351
\bibitem[1994]{martin}Mart\'{\i}n, E.L., Rebolo, R., Magazz\`u, A., \& Pavlenko, Ya.V. 1994, A\&A, 282, 503 (M94)
\bibitem[2006]{montalban}Montalb\`an, J., D'Antona, F. 2006, MNRAS, 370, 1823
\bibitem[1974]{nichiporuk}Nichiporuk, W., Moore, C.B. 1974, Geochim. Cosmochim.
Acta, 38, 1691
\bibitem[1999]{palla99}Palla, F., Stahler, S.W. 1999, ApJ, 525, 772
\bibitem[2000]{palla00}Palla, F., Stahler, S.W. 2000, ApJ, 540, 225
\bibitem[2005]{pallaONC05}Palla, F., Randich, S. Flaccomio, E., \& Pallavicini,
R. 2005, ApJ, 626, L49
\bibitem[2007]{pallaONC07}Palla, F., Randich, S., Pavlenko, Ya. V., Flaccomio, 
E., \& Pallavicini, R. 2007, ApJ, 659, L41
\bibitem[1997]{pavlenko97}Pavlenko, Ya.V. 1997, Ap\&SS, 253, 43
\bibitem[2005]{pavlenko05}Pavlenko, Ya.V. 2001, Astron. Rep., 45, 144
\bibitem[2001]{R01}Randich, S., Pallavicini, R., Meola, G., Stauffer, J.R.,
Balachandran, S. 2001, A\&A, 372, 862
\bibitem[2007]{rebull}Rebull, L.M., Padgett, D., MaCabe, C., et al. 2007,
AAS, 211, 1207
\bibitem[2007]{sacco}Sacco, G., Randich, S., Franciosini, E., Pallavicini, 
R., \& Palla, F. 2007, A\&A, 462, L23
\bibitem[2007]{scelsi}Scelsi, L., Maggio, A., Micela, G., et al.
2007, A\&A, 468, 405
\bibitem[2000]{siess}Siess, L., Dufour, M., \& Forestini, M. 2000, A\&A, 
358, 593
\bibitem[1973]{sneden}Sneden, C.A. 1973, ApJ, 184, 839
\bibitem[1993]{S93} Soderblom, D.R., Stauffer, J.R., Hudon, J.D.,
\& Jones, B.F. 1993, AJSS, 85, 315 
\bibitem[1982]{plateau}Spite, F., \& Spite, M. 1982, A\&A, 115, 357
\bibitem[1998]{stauf98} Stauffer, J. R., Schultz, G., Kirkpatrick,
J. D. 1998, ApJ, 499, L199
\bibitem[1999]{stauf99} Stauffer, J. R., Barrado y Navascu\'es, D.,
Bouvier, J., et al. 1999, ApJ, 527, 219
\bibitem[2000]{staut} Stout-Batalha, N.M., Batalha, C.C., Basri, G.S. 2000,
ApJ, 532, 474
\bibitem[1989]{strom89}Strom, K.M., Wilkin, F.P., Strom, S.E., \& Seaman, R.L. 1989, ApJ, 98, 1444
\bibitem[1988]{walter88}Walter, F.M., Brown, A., Mathieu, R., Myers, P., \& Vrba, F. 1988, AJ, 86, 197
\bibitem[2005]{white05}White, R.J., Hillenbrand, L.A. 2005, ApJ, 621, L65
\bibitem[1996]{wichmann96}Wichmann, R., Krautter, J., Schmitt, J.H.M.M., 
et al. 1996, A\&A, 312, 439
\bibitem[2000]{wichmann00}Wichmann, R., Torres, G., Melo, C.H.F., et al.
2000, A\&A, 359, 181 (W00)
\end{thebibliography}
\end{document}